\documentclass{aastex63}
\usepackage{amsmath}
\usepackage{placeins}
\newcommand\totastcount{4420 }  %
\newcommand\totobscount{82,548 }  %
\newcommand\totnfits{4685 }
\newcommand\nbootstraptrials{200 }
\newcommand\Dperp{1.1645 } %
\newcommand\DperpNeoThis{1.247 } %
\newcommand\DperpNeo{2.336 }
\newcommand\nastsoverlapNEO{2651 }  
\newcommand\nastfitsoverlapNEO{2789 }  
\newcommand\percentDagreementfive{48.7}    
\newcommand\percentDagreementten{74.4}
\newcommand\percentDagreementfifteen{90.5}
\newcommand\percentDagreementtwenty{97.9}
\newcommand\naststax{791 }

\shorttitle{Analysis of four-band WISE observations of asteroids}
\shortauthors{Myhrvold et al.}

\usepackage{hyperref}
\hypersetup{colorlinks=true,citecolor=blue,linkcolor=blue,urlcolor=blue,breaklinks=true}

\begin{document}

\title{Analysis of four-band WISE observations of asteroids}

\correspondingauthor{Nathan Myhrvold}
\email{nathanm@intven.com}

\author[0000-0003-3994-5143]{Nathan Myhrvold}
\affiliation{Intellectual Ventures, 3150 139th Ave SE, Bellevue, WA 98005, USA}

\author[0000-0003-4736-4728]{Pavlo Pinchuk}
\affiliation{Department of Physics and Astronomy, University of California, Los Angeles, CA 90095, USA}

\author[0000-0001-9798-1797]{Jean-Luc Margot}
\affiliation{Department of Earth, Planetary, and Space Sciences, University of California, Los Angeles, CA 90095, USA}
\affiliation{Department of Physics and Astronomy, University of California, Los Angeles, CA 90095, USA}

\begin{abstract} %
We analyzed 82,548 carefully curated observations of 4,420 asteroids
with Wide-field Infrared Survey Explorer (WISE) 4-band data to produce
estimates of diameters and infrared emissivities.  We also used these
diameter values in conjunction with absolute visual magnitudes to
infer estimates of visible-band geometric albedos.  We provide
solutions to 131 asteroids not analyzed by the NEOWISE team and to
1,778 asteroids not analyzed with 4-band data by the NEOWISE team.
Our process differs from the NEOWISE analysis in that it uses an
accurate solar flux, integrates the flux with actual bandpass
responses, obeys Kirchhoff's law, and does not force emissivity values
in all four bands to an arbitrary value of 0.9.  We used a regularized
model-fitting algorithm that yields improved fits to the data.  Our
results more closely match stellar-occultation diameter estimates than
the NEOWISE results by a factor of $\sim$2.  Using 24 high-quality
stellar-occultation results as a benchmark, we found that the median
error of 4-infrared-band diameter estimates in a carefully curated
data set is 9.3\%.  Our results also suggest the presence of a
size-dependent bias in the NEOWISE diameter estimates, which may
pollute estimates of asteroid size distributions and slightly inflate
impact-hazard risk calculations.  For more than 90\% of asteroids in
this sample, the primary source of error on the albedo estimate is the
error on absolute visual magnitude.
\end{abstract}

\section{Introduction} \label{sec:intro}

Accurate asteroid size estimates are important in a number of
contexts.  They yield the size-frequency distribution of asteroids,
which provides critical insights into the collisional evolution of the
main-belt population \citep[e.g.,][]{bott15}.
Sizes are also required to
compute asteroid mass densities,
whose fractional uncertainties scale as the cube of the fractional
uncertainty on size.  Asteroid densities are useful indicators of
composition and internal structure \citep[e.g.,][]{sche15}.
Size and density provide immediate estimates of the strength of the
Yarkovsky orbital drift \citep{gree20}, which has profoundly shaped
the dynamical evolution of the asteroid belt and the delivery of
meteorites to Earth \citep[e.g.,][]{bott06,nesv15,derm21}.  In the impact
hazard context, size and density inform trajectory predictions
\citep[e.g.,][]{farn15} as well as preferred mitigation options
\citep[e.g.,][]{harr15}.  In the exploration and exploitation
contexts, size influences the attractiveness of targets and the range
of possible mission scenarios \citep[e.g.,][]{abel15}.

The best asteroid size measurements are obtained by spacecraft
encounters, Earth-based radar observations with large signal-to-noise
ratios \citep[e.g.,][]{ostr02,benn15}, or stellar occultations with
several chords \citep{hera20}.  These techniques yield tens or
hundreds of size estimates.  Here, we use an infrared technique and
data set that can yield thousands of estimates.  The Wide-field
Infrared Survey Explorer (WISE) mission successfully conducted a
mid-infrared survey of the entire sky at high sensitivity
\citep{Wright2010}.  In the process, it yielded observations for over
100,000 asteroids in four infrared bands (W1--W4) centered at 3.4, 4.6,
12, and 22 $\mu$m.  These observations have the potential to provide a
large number of asteroid size measurements with $\sim$10\% fractional
uncertainties \citep[][and references therein]{main15}.

The NEOWISE team has conducted a pioneering analysis of the WISE
asteroid data set \citep[e.g.,][and subsequent publications\footnote{\href{https://neowise.ipac.caltech.edu/publications.html}{https://neowise.ipac.caltech.edu/publications.html}}]{main11}.
However, questions
remain about the validity of certain assumptions used in this
analysis
and the soundness of some interpretations \citep[e.g.,][]{prav12,hanu15,Myhrvold2018ATM,
  Myhrvold2018Empirical}.
Independent
analyses of this high-quality data set and contribution of open-source
tools \citep[e.g.,][]{moey20}
can help maximize the science return from WISE.
We aim to contribute to this effort with a fully documented
analysis of the WISE asteroid data set.

Our article is organized as follows.  Section~\ref{sec:data} describes
data selection.  Section~\ref{sec:tm} describes thermal modeling
theory and implementation.  \replaced{Results are presented in
  Section~\ref{sec:results}, followed by a discussion and
  conclusions.}{Results are presented in Section~\ref{sec:results},
  followed by a discussion of model limitations in
  Section~\ref{sec-limitations} and conclusions in
  Section~\ref{sec:conclusions}.}

\section{Data Selection} \label{sec:data}
We selected a small, high-quality subset of the data from the WISE
All-Sky Data
Release\footnote{\href{https://wise2.ipac.caltech.edu/docs/release/allsky/}{https://wise2.ipac.caltech.edu/docs/release/allsky/}}.
These data were obtained during the WISE full-cryogenic mission phase,
from 7 January 2010 to 6 August 2010.
We began by querying the WISE All-Sky Single
Exposure Level 1b Source Catalog\footnote{\href{https://irsa.ipac.caltech.edu/}{https://irsa.ipac.caltech.edu/}}
with the ``moving object search'' for all
asteroids listed in the Minor Planet Center MPCORB
database\footnote{\href{https://minorplanetcenter.net/iau/MPCORB.html}{https://minorplanetcenter.net/iau/MPCORB.html}}
(2021/06/08 version).
The queries returned a list of observation dates
recorded as
Modified Julian Date (MJD) values.  We used these MJD values to query
the JPL HORIZONS
service\footnote{\href{https://ssd.jpl.nasa.gov/?horizons}{https://ssd.jpl.nasa.gov/?horizons}}
and obtain the asteroid $H$ and $G$ values, positions, distances to the
Sun and to the WISE spacecraft, and Sun-Target-Observer (phase)
angles.
We queried HORIZONS because the ephemeris values recorded in the WISE
catalog sometimes differ from up-to-date values by a substantial
amount, possibly because the WISE catalog may not incorporate the
latest astrometric data.
HORIZONS also provides distances and phase angles, whereas the WISE moving object search does not.
This process yielded approximately 3.2 million
records.
We then queried the WISE All-Sky Single Exposure Level 1b Source
Catalog and performed a cone search around each one of these positions
with radius 10 arcsec,
and stored these results for subsequent processing.

After this initial data selection, we applied multiple filters to the data.  First, we checked for ``conjunction'' cases, where two or more
\replaced{primary}{presumed-asteroid} sources have an overlapping cone search, i.e., the center of one cone search is within 10 arcsec of another cone search center.  \added{This situation can only occur if the asteroid ephemeris positions are within 10 arcsec of each other.}  In these instances, the WISE catalog unexpectedly reports only one source with the same position for all of the cone search queries.   We found a total of $\sim$12,000
such conjunctions.
We removed these observations from our data set because it is impossible to identify which asteroid is reported and because two or more sources may contribute to the flux detected in a single WISE pixel.  The WISE pixel scales are 2.757 arcsec for W1--W3 and 5.5 arcsec for W4 \citep{Wright2010}.

Second, we eliminated observations that are much more likely to be
stars than asteroids.  Because the flux in W3 and W4 for greybodies
with asteroid temperatures is expected to exceed the flux in W1 and W2
\citep[][Fig.~1]{Myhrvold2018ATM}, we eliminated observations in
all bands at each observation epoch where the signal-to-noise ratio (S/N)
in W3 or W4 was lower than 2.  This step removed 419,039 observations
that might otherwise have eluded the next set of filters.

Next, we eliminated 96,685 observations on the basis of five quality
indicators reported in the WISE
database\footnote{\href{https://wise2.ipac.caltech.edu/docs/release/allsky/expsup/}{https://wise2.ipac.caltech.edu/docs/release/allsky/expsup/}}.
All five quality filters were applied on a per-band basis.  The first
three quality filters are identical to those described by the NEOWISE team.
We discarded observations with nonexistent (``Null'') or poor flux
determinations, i.e., those with photometric quality flags
(``ph\_qual'') $\neq$ ``A'' or ``B''.
Observations susceptible to contamination and confusion, as indicated
by an artifact flag (``cc\_flags'') $\neq 0$, were also discarded.  We
also removed any observation with a low photometric signal-to-noise
ratio (S/N), as indicated by a ``w$k$snr'' ($1 \leq k \leq 4$) field
value $< 4$.  The remaining filters may be unique to this work as they
are not described in NEOWISE team publications.  We removed any
observations with poor point spread function (PSF) fits, specifically
those with reduced chi-squared value $> 2$, as reported by the
``w$k$rchi2'' ($1 \leq k \leq 4$) fields.  This filter is consistent
with
suggestions in the WISE Explanatory
Supplement\footnote{\href{https://wise2.ipac.caltech.edu/docs/release/allsky/expsup/}{https://wise2.ipac.caltech.edu/docs/release/allsky/expsup/}}
that
photometry may be
degraded in the case of poor PSF fits with
``w$k$rchi2'' $>$2.  Finally, we removed saturated observations where
the saturation indicator ``w$k$sat'' ($1 \leq k \leq 4$) was different
from 0, although most of these observations had already been
eliminated by the PSF filter.

The fourth filter was designed to handle cases where there is
potential confusion
between asteroids
and other astronomical sources.  To do this, we used the AllWISE
(co-added) Source
Catalog\footnote{\href{https://wise2.ipac.caltech.edu/docs/release/allwise/}{https://wise2.ipac.caltech.edu/docs/release/allwise/}}.
The requirement for inclusion in this catalog is that a source must
have been detected by WISE in at least 3 observations at a photometric
S/N $\geq 4$.
Because the sources are approximately stationary on the sky, they
{\em cannot} be attributed to asteroids, and we refer to them as
``background.''  Conversely, the “foreground” sources
found with
the WISE All-Sky Single Exposure Level 1b Source Catalog
could be an asteroid, or a background source, or a mix.  We
obtained a list of potentially confusing background sources by
performing a cone search with a radius of 10 arcsec of the ephemeris
position corresponding to every asteroid observation that remained
after application of our first
three filters.
There were approximately 603,000 cases where
the combined results of a moving-object search and a background-object search at a given ephemeris position 
returned one or more foreground sources and at least one
background source.
The first step towards resolution of the potential confusion 
was to determine possible matches between the sources.
We calculated the angular distance between each foreground and
background source.  
If each band in a foreground source $F$ was found in
a background source $B$, and $B$ was the closest background source to $F$,
and the distance between them
was $\leq 2.60$ arcsec (a 95\% quantile threshold described in the next paragraph),
then $F$ and $B$ were considered to be matched, and both sources were
removed from further consideration.
This source-matching filter eliminated approximately 13,000
observations and resolved approximately 300 other cases where there
was only a single foreground source remaining after the elimination.
However,
589,124 cases were still in need of resolution,
each of which
had
multiple foreground and/or background sources.  The second step
towards resolution was to eliminate observations in bands where
confusion was possible.
If a measurable flux from an unmatched background source was detected in any band, then observations from any foreground source were eliminated in that band.   
This approach does not consider other quantities,
such as distance between foreground and background sources, or
asteroid flux values in the shared bands.  Although one could
certainly develop methods that include these quantities, we chose a
simpler algorithm for our first analysis of these data.

The fifth filter was based on the distance between observed WISE
position and HORIZONS-predicted ephemeris position.
To calculate suitable distance thresholds, we assembled a subset of
all available observations.  Approximately 1.7 million observations
meet the following three conditions: (1) a single primary foreground
source is returned per ephemeris position in the WISE All-Sky Single
Exposure Level 1b Source Catalog query; (2) no background sources are
detected within 10 arcsec of that ephemeris position in the AllWISE
(co-added) Source Catalog query; (3) the observation passes the
S/N, photometric quality, artifact, and PSF quality filters described
above.  This subset of primary sources is least likely to be affected
by background contamination and thus most likely to be attributable to
asteroids.
With this subset, we calculated probability density curves of the
distance between the WISE-reported source positions and the
HORIZONS-predicted ephemeris positions. We used the 95\% quantiles as
a maximum allowable angular distance threshold.  While this choice of
threshold undoubtedly removed some (up to 5\%) valid data points, it
also greatly reduced contamination with spurious nonasteroid sources.
Overall, 95\% of the presumed-asteroidal sources were within 2.60 arcsec of the
HORIZONS-predicted ephemeris position.  Upon closer examination, the
distributions of distances to ephemeris position have different
characteristics for numbered and unnumbered asteroids, as well as a
function of the lowest WISE band in which the source is detected.
Table \ref{tab-dist} describes the adopted thresholds in each case.  
The distance between the calculated ephemeris position and the
position reported by WISE is a mixture of uncertainties of different
origin.  These uncertainties include errors in our knowledge of the
orbital parameters of the asteroids and nongravitational
influences,
but also astrometric errors resulting from
PSF fits in the WISE pipeline, which depend in part on the size of the
WISE detector pixels and S/N.
A single foreground source must be below the thresholds in Table \ref{tab-dist} to
be retained.  In the case where two or more foreground objects
survived background matching (fourth filter), the closest source was
chosen so long as (1) its distance to the ephemeris position was
below the thresholds in Table \ref{tab-dist} and (2) the next closest foreground
source was at least 6 arcsec from the ephemeris position.

\begin{table}[h]
  \begin{centering}
  \begin{tabular}{llll}
                   & Lowest band is W1,2 & Lowest band is W3 & Lowest band is W4 \\
    \hline
    Numbered       & 1                   & 2.58              & 5.14\\
    Unnumbered     & 2.26                & 5.83              & 10 (no filter)\\
  \end{tabular}
  \caption{Adopted angular distance thresholds in arcseconds for numbered and unnumbered asteroids, as a function of the lowest WISE band in which the source is detected with S/N $\geq$ 4.
  Observations where the distance between reported position and ephemeris prediction exceeded the relevant threshold were discarded.}
  \label{tab-dist}
  \end{centering}
\end{table}

\begin{figure}[htb]
  \begin{center}
    \includegraphics[width=5in]{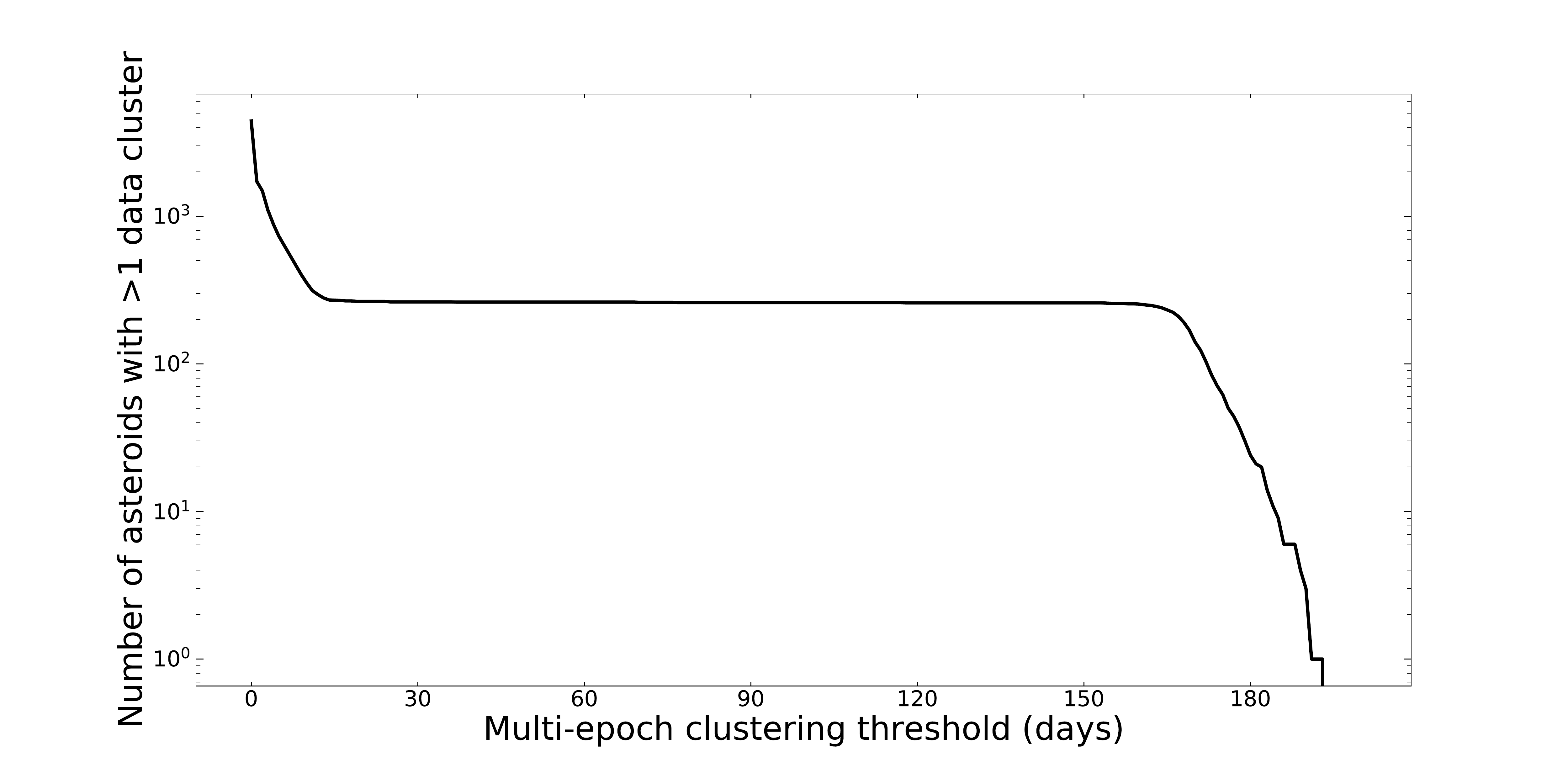}
\caption{Number of asteroids affected by a multi-epoch clustering threshold value of $N$ days as a function of $N$.  Note that the curve flattens at a threshold value of $\sim$30 days.}    
\label{fig:cluster_hist}
\end{center}
  \end{figure}

In the
sixth filter, we applied a simple clustering algorithm to split the
data for each asteroid into one or more subsets of observation epochs.
Specifically, we split the data set whenever we found that consecutive
observations were more than 30 days apart
(Figure~\ref{fig:cluster_hist}).  
As demonstrated in Figure~\ref{fig:cluster_hist}, the time elapsed between observation clusters often exceeds 30 days, which is due to the interaction between the WISE observational cadence and the asteroid positions.  
We
calculated independent solutions for each cluster in part to
facilitate comparisons with results from the NEOWISE team, which
presented different diameter solutions for different epochs.
The separate 
solutions 
also help manage large differences in phase angle across observations.

Seventh, we removed any subset of data that did not include at least
three remaining observations in each of the four WISE bands.  The
three data points are required to enable fits with three adjustable
parameters in each band (Section~\ref{subsec:reg_fitting}).

Finally, we implemented a basic pre-fit outlier rejection procedure.
Any observation with a flux $\geq$ 1.5 mag from the next lowest or
highest flux within its cluster of observational epochs was removed.
This step removed a very small number of observations and did not
change the number of asteroids in the filtered data set.

After application of the data filters and outlier rejection, we were
left with
\totobscount observations grouped into \totnfits valid clusters of
observations and representing \totastcount asteroids.  Except for 16
asteroids that have provisional designations at the time of writing,
all asteroids have received a permanent IAU number.  \added{The median number
of observations per asteroid is 8, 11, 13, and 15 in W1, W2, W3, and W4, respectively.}

\section{Thermal Modeling} \label{sec:tm} 

\subsection{NEATM} \label{subsec:neatm} 
\begin{figure}[htb]
  \begin{center}
    \includegraphics[width=3in]{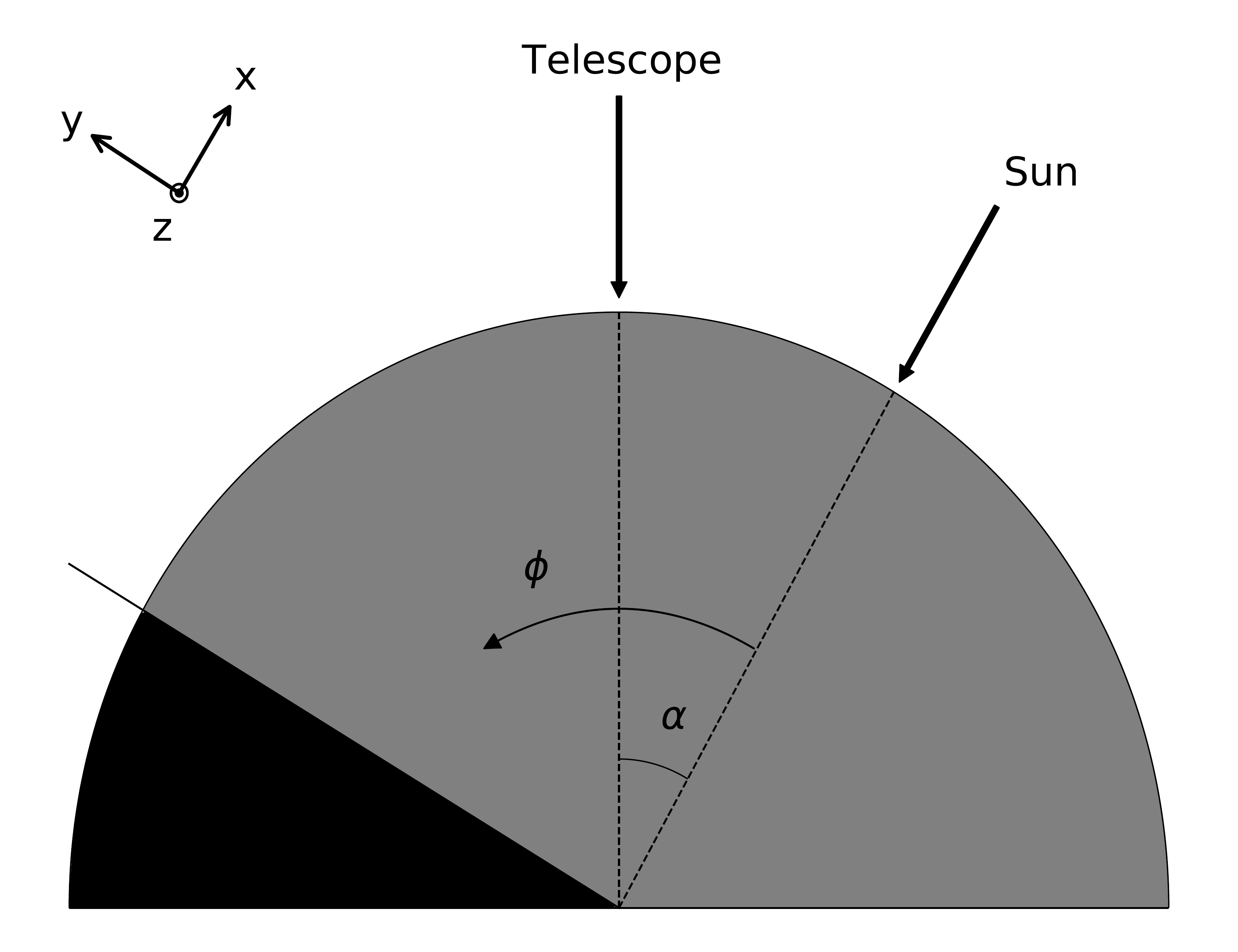}
\caption{Definition of coordinate frame for NEATM calculations.  Both the Sun and telescope are in the plane of the page.  The spin vector is not shown.}
\label{fig:neatm}
\end{center}
  \end{figure}

We fit the asteroid thermal data using a reparameterized version of
the Near-Earth Asteroid Thermal Model (NEATM).  
For these calculations, we define a right-handed frame centered on the
asteroid with $\vec{x}$ pointing at the Sun, $\vec{z}$ perpendicular
to the plane that contains $\vec{x}$ and the observer's line of sight,
and $\vec{y}$ completing the frame (Figure~\ref{fig:neatm}).  We then
define the usual spherical coordinate system with longitude $\phi$
measured from $\vec{x}$ in the $xy$ plane and colatitude $\theta$
measured from the positive $\vec{z}$ axis. 
The standard version of NEATM was originally derived by
\citet{Harris98} and gives the thermal flux of an asteroid with
diameter $D$ and emissivity $\epsilon(\lambda)$
for an asteroid-observer distance $r_{\rm ao}$
and phase angle $\alpha$ as 
\begin{equation}\label{eq:NEATM_flux_conv}
  F_{\rm NEATM}(\alpha, \lambda) = \frac{\epsilon(\lambda) D^2}{4 {r_{\rm ao}}^2}  \int_{\theta=0}^{\theta=\pi} \int_{\phi=\alpha-\pi / 2}^{\phi=\alpha+\pi / 2} B(\lambda, T(\theta, \phi)) \sin^2 \theta \cos(\phi - \alpha)\, d\phi\, d\theta,
\end{equation}
where $B(\lambda, T)$ is the Planck distribution given by 
\begin{equation}\label{eq:plank}
B(\lambda, T) = \frac{2 h c^2}{\lambda^5}\frac{1}{\exp(\frac{h c}{\lambda k T}) - 1},
\end{equation}
and $T(\theta, \phi)$ is the temperature distribution given by
\begin{equation}\label{eq:T_dist_conv}
    T(\theta, \phi) = T_{\rm ss}\max{(0, \sin\theta\cos\phi)}^{1/4},
\end{equation}
where $T_{\rm ss}$ is the sub-solar temperature.  In these
expressions, $\lambda$ is the wavelength, $h$ is Planck's constant,
$k$ is Boltzmann's constant, and $c$ is the speed of light.  The
integration domain includes the hemisphere visible to the observer
with all nightside ($|\phi| > \pi/2$) emission set to zero through
equation (\ref{eq:T_dist_conv}).  The prescription for zero emission
on the nightside is a feature of NEATM.  Appendix~\ref{app:neatm}
describes an equivalent NEATM formulation with a different choice of angles.

The sub-solar temperature $T_{\rm ss}$ is given by
\begin{equation}\label{eq:T_ss}
T_{\rm ss} = \left(\frac{S_0 (1 - A)}{\epsilon \sigma \eta {r_{\rm as}}^2}\right)^{1/4}.
\end{equation}
where $S_0 = 1360.8$ W/m$^2$ is the solar constant, 
$A$ is the Bond albedo, $\sigma$
is the Stefan-Boltzmann constant, $r_{\rm as}$ is the asteroid-Sun distance,
and $\eta$ is a free
parameter in NEATM referred to as 
the ``beaming parameter''.
The Bond albedo is typically approximated
as the visible-band Bond
albedo \citep{Hapke2012}
\begin{equation}\label{eq:av}
A \approx A_V = p_{\rm V} q,
\end{equation}
where $p_{\rm V}$ is the visible-band geometric albedo and $q$ is the
visible-band phase integral, which is typically obtained by
integrating the empirical
asteroid phase function $\psi_{\rm HG}(\alpha, G)$
\citep{Bowell89}. Integration of $\psi_{\rm HG}$ over all phase angles gives $q$ as
\begin{equation}\label{eq:q_G}
q = 0.285596 + 0.656288 \times G, %
\end{equation}
which is
different from the frequently used but incorrect expression described by \citet{Bowell89}:
\begin{equation}\label{eq:q_G_old}
q = 0.290 + 0.684 \times G. %
\end{equation}
For a typical value of the slope parameter G=0.15, the two expressions differ by 2\%.

\citet{Myhrvold2018ATM} pointed out that two of the fitting parameters in the traditional NEATM,
$p_{\rm V}$ and $\eta$,
appear in the numerator and denominator of Equation (\ref{eq:T_ss}),
thereby increasing the likelihood of numerical instabilities or
convergence to local minima.
To remedy this issue, \citet{Myhrvold2018ATM} proposed a reparameterized version of 
NEATM, which introduced a new fitting  parameter, the pseudo-temperature $T_1$:
\begin{equation}\label{eq:T1}
T_1 = \left(\frac{S_0 (1 - A)}{\epsilon \sigma \eta} \right)^{1/4}.
\end{equation}
This parameter has the physical interpretation of $T_1 = T_{\rm ss}$ at 
$r_{\rm as} = 1$. With this reparameterization, equation (\ref{eq:T_ss}) becomes
\begin{equation}\label{eq:Tss_T1}
T_{\rm ss} = \frac{T_1}{\sqrt{r_{\rm as}}}.
\end{equation}
When fitting thermal flux with the reparameterized version of NEATM to
asteroid data, the only
floating parameters
are $T_1$, which controls the shape of the curve, and the asteroid
diameter $D$, which controls the amplitude of the curve.  Importantly,
the reparameterization highlights the fact that there are only two, not
three, degrees of freedom in this problem \citep{Myhrvold2018ATM}.
We emphasize that fitting thermal flux does not rely in any way on the
absolute visual magnitude $H$ of the asteroid.

\subsection{Reflected Sunlight} \label{subsec:rsun} 
\begin{figure}[htb]
  \begin{center}
    \includegraphics[width=5in]{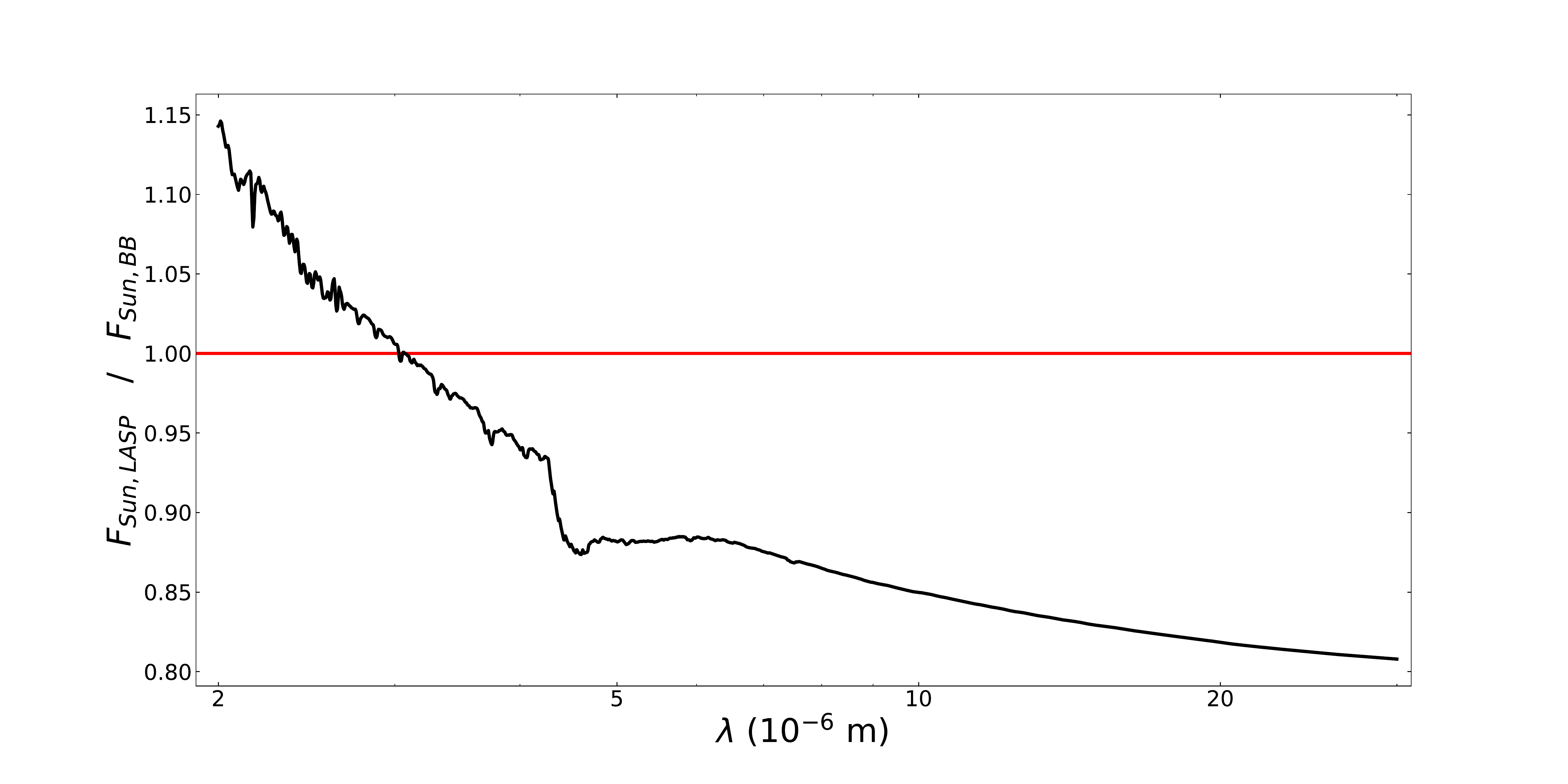}
\caption{Ratio of the 2010 yearly averaged solar flux from satellite
  measurements to the flux of a blackbody at 5778 K. Deviations
  of up to 20\% occur at WISE operating wavelengths.}
\label{fig:solar_spec_compare}
\end{center}
  \end{figure}

Because 
WISE observations of asteroids are sensitive to
reflected
sunlight, we add a reflected
flux component to the model
\begin{equation}\label{eq:RFL_flux}
F_{\rm REFL}(\alpha, \lambda) = p(\lambda) \frac{D^2}{4 {r_{\rm ao}}^2} \psi_{\rm HG}(\alpha, G)  F_{\rm Sun}(\lambda),
\end{equation}
where $p(\lambda)$ is the geometric albedo
and $F_{\rm Sun}(\lambda)$ is the solar flux incident on the asteroid.
The solar flux is often estimated with a blackbody model at 5778 K
\begin{equation}\label{eq:Sun_flux_inc}
F_{\rm Sun, \rm BB}(\lambda) = \frac{\pi {R_{\rm Sun}}^2}{{r_{\rm as}}^2} B(\lambda, 5778\, {\rm K}),
\end{equation}
where $R_{\rm Sun}$ is the solar radius.
However, the true solar spectrum deviates by up to 20\% from the
blackbody model at near-infrared wavelengths (Figure
\ref{fig:solar_spec_compare}).  In order to provide the best possible
model of reflected light, we used an actual solar spectrum obtained
with satellite measurements.  Specifically, we used the 2010 yearly
averaged solar
spectrum\footnote{\href{http://lasp.colorado.edu/lisird/data/nrl2\_ssi\_P1Y/}{http://lasp.colorado.edu/lisird/data/nrl2\_ssi\_P1Y/}}
from the Solar Radiation and Climate Experiment (SORCE)
compiled by the University of Colorado’s Laboratory for
Atmospheric and Space Physics (LASP) and the Naval Research Laboratory
\citep{LASP}, i.e., we set $F_{\rm Sun} = F_{\rm Sun, \rm LASP}$.

\subsection{Integration over WISE Bands} \label{subsec:RSR}
The total flux we expect to observe from a given asteroid is the 
sum of the emitted and reflected components:
\begin{equation}\label{eq:tot_flux}
  F_{\rm total}(\alpha, \lambda) = F_{\rm NEATM}(\alpha, \lambda) + F_{\rm REFL}(\alpha, \lambda).
\end{equation}
Equation \ref{eq:tot_flux} gives the total model flux at a single 
wavelength. However, the WISE bands span bandwidths on the order of 
$\sim2-10$ microns \citep[e.g.,][Figures 6 and 7]{Wright2010}. 
In order to determine the flux measured by the receiver, we integrate 
Equation \ref{eq:tot_flux} over the receiver response:
\begin{equation}\label{eq:RSR_int}
  \langle F_{\rm total} \rangle = \frac{\int \lambda R(\lambda) F_{\rm total}(\alpha, \lambda) \, d\lambda}{\int \lambda R(\lambda) \, d\lambda},
\end{equation}
where $R(\lambda)$ is the relative spectral response (RSR) of the WISE 
sensors,\footnote{\href{https://wise2.ipac.caltech.edu/docs/release/allsky/expsup/sec4\_4h.html}{https://wise2.ipac.caltech.edu/docs/release/allsky/expsup/sec4\_4h.html}}
measured in terms of quantum efficiency per photon.
A separate integration is performed for each WISE band W$_i$.
When integrating over the W4 RSR, we scaled the wavelengths by a factor 
of 1.033 as recommended by \citet{brown2014}, who found that this correction
reduced the residual between the measured W4 photometry and the photometry
synthesised from spectra of galaxies and planetary nebulae.

In order to avoid redundant computations when fitting thousands of
models, we calculated Equation \ref{eq:RSR_int} for a fine grid of
$T_{\rm ss}$ and $\alpha$ values, with the integrand of Equation
\ref{eq:NEATM_flux_conv} as the model flux. This allowed us to replace
the computationally expensive integrations with efficient table
lookups and bilinear interpolations between table entries, which
considerably reduced the processing time.
In the most relevant temperature range of $T_{\rm ss} = 200-400$K, the
fractional errors introduced by the table lookup procedure are $<$1
ppm.  The worst possible fractional errors occur for extremely low
temperatures, $T_{\rm ss} = 20$K, and remain $<$0.5\%.

Before comparing the model flux to the observations, we converted the calculated flux to units 
of magnitude in order to match the \replaced{fluxed}{flux} reported by WISE:
\begin{equation}\label{eq:mag_convert}
  F_{\rm model} = -2.5 \,\log_{10}\langle F_{\rm total} \rangle.
\end{equation}

\subsection{Implementation of Kirchhoff's Law} \label{subsec:Kirchhoff} 
Emissivity and albedo are related through Kirchhoff's law of thermal radiation, which can
be expressed as follows:
\begin{equation}\label{eq:Kirchhoffs_law}
p(\lambda) = \frac{1 - \epsilon(\lambda)}{q}.
\end{equation}
When fitting WISE data, we allowed for one $\epsilon(\lambda)$ value 
per WISE band, such that 
\begin{equation}\label{eq:WISE_eps}
\epsilon(\lambda) = 
\begin{cases}
    \epsilon_{1}, & \lambda \in W1\\
    \epsilon_{2}, & \lambda \in W2\\
    \epsilon_{3}, & \lambda \in W3\\
    \epsilon_{4}, & \lambda \in W4
\end{cases}
\end{equation}
Because W3 and W4 are dominated by thermally emitted light, the
emissivities for those bands are typically set (somewhat arbitrarily)
by some investigators to $\epsilon_{3} = \epsilon_{4} = 0.9$ during
fitting.
This choice is reasonable for many asteroids and has widely been used in past studies.
However, other
values in the range
$\sim0.5 < \epsilon_i < 1$
are physically reasonable both for likely asteroid minerals and surface conditions.
We allow for this possibility with a regularized modeling approach
(Section~\ref{subsec:reg_fitting}).  We found that
relaxing the arbitrary constraints on emissivities
was important
in order to obtain satisfactory fits to the data.

\subsection{Least-Squares Model Fits} \label{subsec:tm_fitting}
During the fitting procedure, we compared the model flux to the
observed flux $F_{\rm obs}$ and adjusted model parameters to minimize
an error term $L$. A common approach is to minimize the chi-squared 
error $L = \chi^2$ using a least-squares fitting routine. 
The $\chi^2$ error is calculated as
\begin{equation}\label{eq:chi_sq_err}
  \chi^2 = \sum_i \left(\frac{F_{{\rm obs}, i} - F_{{\rm model}, i}}{\sigma_i} \right)^2,
\end{equation}
where the indices $i$ represent individual data points for a given
asteroid and $\sigma_i$ is the measurement error for the $i^{th}$
observation.
The summation in Equation \ref{eq:chi_sq_err} is performed over all                                               
observations from all WISE bands
within a subset of observation epochs as defined in
Section~\ref{sec:data}.  

The chi-squared minimization is common in asteroid thermal modeling,
but it is problematic with WISE data because the observation error
estimates produced by the WISE pipeline are problematic
\citep{hanu15,Myhrvold2018ATM, Myhrvold2018Empirical}.  It is also
problematic with the NEATM model, which relies on the assumption of
spherical asteroids.  In a rigorous chi-squared implementation, the
error term in
Equation (\ref{eq:chi_sq_err}) must include modeling error, e.g.,
deviations from a constant surface area due to nonspherical asteroid
shapes.  Using a chi-squared approach with incorrect error values can
trap the solution in false minima \citep{ratkowsky1983}.

\subsection{Regularized Model Fits} \label{subsec:reg_fitting} 

\begin{figure}[h!]
\plotone{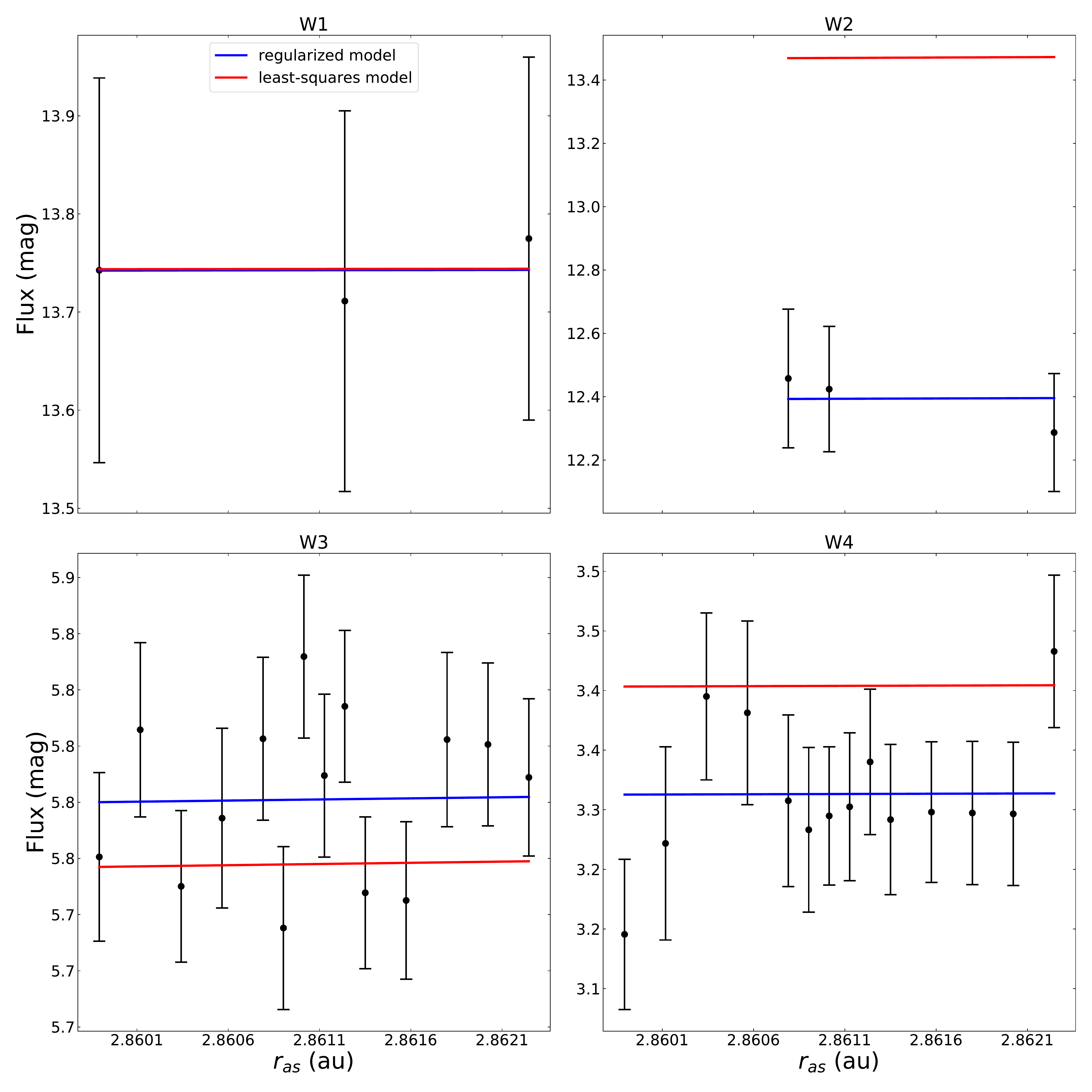}
  \caption{Best-fit models to 
   38
  observations of asteroid 
  7451 Verbitskaya 
  with the standard NEATM (least-squares model) and our regularized approach (regularized model). Flux values shown in this figure are displayed as a function of asteroid-Sun distance and reduced to the values that would be observed at an asteroid-observer distance $r_{\rm ao}$ = 1 au.
  The least-squares model completely misses the data in W2, and the miss is not due to the small number of points in W2 because the least-squares model fits the data well in W1 with the same number of points.  In addition, the small number of points in W2 does not prevent the regularized model from fitting the data.  The $\chi^2$ values are 
    152 (least-squares model) and 25 (regularized model) whereas the sums of squares of (unweighted) residuals are 3.7 (least-squares model) and 0.075 (regularized model). 
  \label{fig:poor_fit}}
\end{figure}

\citet{Myhrvold2018Empirical} first demonstrated that
model fits performed as described in Section \ref{subsec:tm_fitting}
can result in poor fits to the data (see his Figure 3).
Specifically, he
showed that
many (up to 50\%) NEOWISE 
model fits miss the data completely in one or more WISE bands.
In a \replaced{preprint}{withdrawn manuscript}, \citet{wrig18} ascribed the problem of fits missing the data in the case of two asteroids to a previously unreported software bug in the NEOWISE modeling code that was identified and fixed in 2011.  This bug potentially corrupts all NEOWISE model estimates obtained prior to the bug fix.  While the bug may explain the problem with reported NEOWISE results for these two asteroids, the problem of model fits missing data is widespread and points to a more fundamental issue.
Indeed, a cursory analysis suggests that at least $\sim1600$ asteroids out of
\totastcount ($\sim36\%$) in our sample
exhibit least-squares model fits with unacceptably low quality.
We obtained this number by counting the number of asteroids that, in
at least one of the bands, have a residual larger than the measured
error for every point, and the residuals in that band are either all
positive or all negative.
Figure \ref{fig:poor_fit} shows an example.

We found that the quality of the
model fits can be
improved by relaxing the restrictions placed on the per-band
emissivity values $\epsilon_i$,
which are arbitrarily set to 0.9 in the NEOWISE solutions.
Specifically, we allowed all four
values to vary in the range $0 < \epsilon_i < 1$,
although results rarely yield
$\epsilon_i < 0.7$.  However, the additional degrees of freedom
introduced with varying emissivities are not handled well by
conventional least-squares curve fitting.  Therefore, we used a
regularized method and
reined in
the additional freedom in the model with a
regularization term that helps satisfy two objectives.
First, we wish to favor models with high emissivity values.  Second,
we wish to disfavor models with a large dispersion in emissivity
values.  Both of these objectives are aligned with the classic
$\epsilon=0.9$ assumption across all bands and with observational data
\citep[e.g., Figure 1 of][]{Myhrvold2018ATM}.
Laboratory spectra from meteorites and previous work on IR spectra of
asteroids indicate that asteroid materials generally have emissivity
values that are close to unity and do not exhibit substantial
wavelength dependence \citep[Supplementary Material of][and references therein]{Myhrvold2018ATM}.

The regularized model loss is given by the sum of a classic loss function and a regularization term:
\begin{equation}\label{eq:reg_loss}
L  = L_{\rm data} + R_{\rm model}.
\end{equation}
In order to assign approximately equal weights to the classic loss
function and the regularization term, we normalized each one to a
value near unity with the following expressions.
The scaled residual loss $L_{\rm data}$ is given by
\begin{equation}\label{eq:L_data}
L_{\rm data}  = \frac{L_{2} - L_{\rm min}}{L_{\rm min}},
\end{equation}
where $L_{2}$ is the sum of squares of (unweighted) residuals,
$L_{2} = \sum\limits_i (F_{\rm obs, i} - F_{\rm model, i})^2$, and
$L_{\rm min}$ is a lower bound on $L_2$, which we obtained by fitting
the model (Equation \ref{eq:tot_flux}) to each individual band W1--W4
with three floating parameters ($T_1$, $D$, and $\epsilon_i$), then
summing the sums of squares of residuals over all four bands.  We
chose $L_2$ instead of $\chi^2$ because of the problems associated with 
 the $\chi^2$ metric described in Section \ref{subsec:tm_fitting}.

The scaled regularization loss is given by
\begin{equation}\label{eq:R_model}
  R_{\rm model} = ||\epsilon - 0.9||.
\end{equation}
where $||\vec{a}|| = \left[ \sum_i a_i^2 \right]^{0.5}$ is the norm
operator and $\epsilon$ is the emissivity vector $\{\epsilon_{1},
\epsilon_{2}, \epsilon_{3}, \epsilon_{4}\}$.  We also obtained fits
with a few other forms of the regularization loss
(Appendix~\ref{app:alternate}), which confirmed the robustness of our
general conclusions but did not perform quite as well in terms of
diameter accuracy.

We used the Limited-memory Broyden--Fletcher--Goldfarb--Shanno algorithm 
with Box constraints \citep[L--BFGS--B;][]{Byrd1995}
to minimize the loss $L$ for every set of observations.
We executed the fitting routine seven times per asteroid with seven
different initial values for $\epsilon_1=\epsilon_2=\epsilon_3=\epsilon_4$ selected from the set
$\{0.6, 0.7, 0.8, 0.85, 0.9, 0.95, 0.99\}$.
In some cases, we detected a loss $L_2$ during the fit
that was lower than our initial estimate of $L_{\rm min}$.
When this situation arose, we updated the value of $L_{\rm min}$
accordingly and restarted the fitting procedure.

\begin{figure}[htbp]
  \begin{center}
    \includegraphics[width=6.5in]{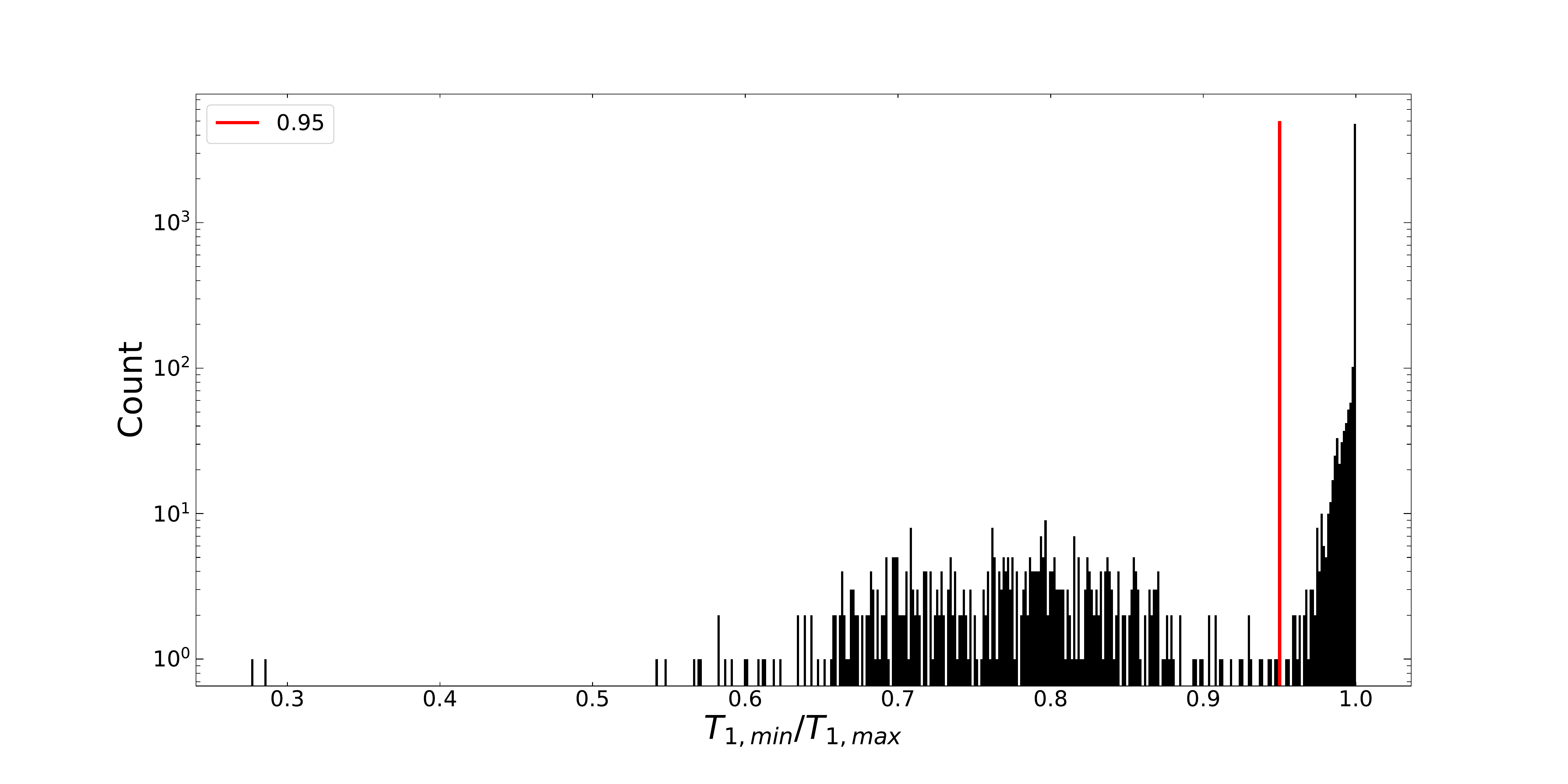}
\caption{Ratio of the largest and smallest $T_1$ values amongst the seven 
solutions to our fitting procedure. For ratios below 0.95, the fits fall into
one of two local minima.
We report the best-fit solution from the high-$T_1$ group.}
\label{fig:T1_ratios}
\end{center}
  \end{figure}

After applying this procedure, we noticed that,
in approximately 8\% of the cases, the seven solutions to our
fitting procedure did not have a single minimum but rather fell into
one of two local minima.  One of the minima occurs at $T_1\sim 230-380$~K
with a comparatively larger diameter $D$, whereas the other minimum
occurs at a larger value of $T_1$ $\sim 330-430$~K but smaller
diameter $D$ (Figure \ref{fig:T1_ratios}).  Because the $T_1$ parameter has
the physical interpretation of subsolar temperature, i.e., $T_1 =
T_{\rm ss}$ at $r_{\rm as} = 1$ (Equation \ref{eq:T1}), we expect most
thermal model fits to have $T_1 = 350-450$~K.  Thus, when two groups of
local minima were detected, we chose to report the best-fit solution
among the high-$T_1$ group.  Specifically, we selected all solutions
with $T_1 \geq 95$\% of the largest $T_1$ in any given fit, and then
selected the best-fit solution (smallest $L$) amongst those.

\subsection{Bootstrap Trials} \label{subsec:bootstrap}
Because the WISE uncertainties are in doubt
\citep{hanu15,Myhrvold2018Empirical} and do not include lightcurve errors,
we turned instead to the Bootstrap method to estimate errors in
diameter and other parameters\citep[e.g.,][]{feigelson2012,ivezic2014}.
We performed \nbootstraptrials bootstrap trials for every asteroid in
order to reduce the influence of any remaining data outliers and to
obtain an estimate of the uncertainty on the model fit parameters.
For each trial, we randomly sampled the asteroid data with replacement
until the number of samples matched the original number of samples.
We repeated the sampling procedure as necessary until the requirement
of at least 3 data points in each of the four WISE bands was met.  We
then performed a model fit (Section \ref{subsec:reg_fitting}) on the
sampled data. We repeated this process \nbootstraptrials times per
asteroid to obtain \nbootstraptrials unique model fits. The final
model parameters as well as their uncertainties were obtained by
taking the mean value and the standard deviations of the
\nbootstraptrials bootstrap fits, respectively.

\section{Results}\label{sec:results}
\subsection{Diameters}
We present \totnfits thermal model fit results for a total of \totastcount asteroids (4326 MBAs and 94 NEAs).
The full table of results is available
\href{https://ucla.box.com/s/4i2at438pvvh92wl8grh9z0ntj071moa}{online}.
A sample of the table is shown in Table \ref{tbl:results}.

\begin{longrotatetable}
\begin{deluxetable}{rrrrrrrrrrrrrrrrrrrr}%
\tablecaption{ Results of thermal model fits showing the asteroid
  number or provisional designation, the unique 7+ digit Spacecraft
  and Planet Kernel (SPK) ID of the Navigation and Ancillary
  Information Facility (NAIF) node of the NASA Planetary Data
  System,
  the mean observation date of each cluster of observational epochs
  and the number of data points for the subset of observations used in
  the fit, and best-fit estimates and errors of the diameter $D$,
  pseudo-temperature $T_1$, emissivities $\epsilon_i$, and geometric
  albedo $p_{\rm V}$.  Parameter uncertainties (1$\sigma$) are obtained by
  calculating the standard deviations of \nbootstraptrials bootstrap
  resampling solutions.  Albedo values are calculated on the basis of
  $H$ values reported by HORIZONS,
and the 
corresponding error values are obtained by assuming an uncertainty of 0.25 mag on 
the $H$ values \citep[][Figure 5]{veres2015}.
The values $p_{\rm V}'$ and $\sigma_{p_{\rm V}'}$ represent the albedo and its error 
calculated using $H$ values from \citet{veres2015} when available (defaulting to 
the
HORIZONS $H$ values otherwise). 
(This table is available in its entirety in a
\href{https://ucla.box.com/s/4i2at438pvvh92wl8grh9z0ntj071moa}{machine-readable form}
in
the online journal. A portion is shown here for guidance regarding
its form and content. \label{tbl:results}}
  \tablehead{
  \colhead{Object} & \colhead{SPK ID} & \colhead{Mean MJD} & \colhead{$N$} & 
  \colhead{$D$ (km)} & \colhead{$\sigma_{D}$} &  
  \colhead{$T_1$} & \colhead{$\sigma_{T_1}$} & 
  \colhead{$\epsilon_1$} & \colhead{$\sigma_{\epsilon_1}$} & 
  \colhead{$\epsilon_2$} & \colhead{$\sigma_{\epsilon_2}$} &  
  \colhead{$\epsilon_3$} & \colhead{$\sigma_{\epsilon_3}$} &  
  \colhead{$\epsilon_4$} & \colhead{$\sigma_{\epsilon_4}$} &  
  \colhead{$p_{\rm V}$} & \colhead{$\sigma_{p_{\rm V}}$} &
  \colhead{$p_{\rm V}'$} & \colhead{$\sigma_{p_{\rm V}'}$} 
  }
\startdata
131 & 2000131 & 55322.809852 & 60 & 30.59 & 0.6 & 405.15 & 2.6 & 0.899 & 0.005 & 0.897 & 0.003 & 0.859 & 0.037 & 0.926 & 0.022 & 0.194 & 0.045 & 0.153 & 0.036 \\
155 & 2000155 & 55395.529229 & 53 & 44.65 & 2.6 & 379.44 & 9.9 & 0.981 & 0.002 & 0.939 & 0.013 & 0.890 & 0.037 & 0.918 & 0.026 & 0.029 & 0.007 & 0.023 & 0.006 \\
170 & 2000170 & 55404.853180 & 42 & 35.42 & 1.8 & 389.20 & 8.2 & 0.856 & 0.012 & 0.861 & 0.018 & 0.849 & 0.018 & 0.917 & 0.014 & 0.256 & 0.065 & 0.202 & 0.051 \\
180 & 2000180 & 55319.783820 & 43 & 24.79 & 2.2 & 396.42 & 14.7 & 0.861 & 0.026 & 0.867 & 0.028 & 0.896 & 0.029 & 0.913 & 0.039 & 0.239 & 0.070 & 0.188 & 0.055 \\
183 & 2000183 & 55234.369656 & 64 & 30.73 & 2.9 & 412.37 & 15.6 & 0.842 & 0.028 & 0.909 & 0.023 & 0.867 & 0.091 & 0.947 & 0.055 & 0.278 & 0.082 & 0.219 & 0.065 \\
183 & 2000183 & 55398.940284 & 40 & 30.76 & 5.2 & 405.50 & 29.0 & 0.850 & 0.053 & 0.887 & 0.029 & 0.862 & 0.094 & 0.944 & 0.054 & 0.277 & 0.114 & 0.218 & 0.090 \\
193 & 2000193 & 55250.762781 & 33 & 28.82 & 2.7 & 374.72 & 16.2 & 0.891 & 0.022 & 0.913 & 0.018 & 0.895 & 0.035 & 0.908 & 0.029 & 0.249 & 0.073 & 0.196 & 0.058 \\
214 & 2000214 & 55246.883604 & 41 & 29.72 & 7.1 & 372.48 & 88.5 & 0.808 & 0.143 & 0.692 & 0.300 & 0.841 & 0.195 & 0.863 & 0.064 & 0.388 & 0.205 & 0.306 & 0.162 \\
239 & 2000239 & 55368.131442 & 44 & 42.88 & 2.8 & 391.03 & 12.5 & 0.968 & 0.005 & 0.955 & 0.011 & 0.902 & 0.025 & 0.907 & 0.020 & 0.060 & 0.016 & 0.047 & 0.013 \\
242 & 2000242 & 55408.924526 & 49 & 46.76 & 3.1 & 353.12 & 13.1 & 0.914 & 0.014 & 0.901 & 0.014 & 0.891 & 0.031 & 0.907 & 0.018 & 0.144 & 0.038 & 0.114 & 0.030 \\
244 & 2000244 & 55208.721440 & 39 & 10.93 & 0.5 & 393.35 & 5.6 & 0.848 & 0.015 & 0.901 & 0.002 & 0.853 & 0.040 & 0.920 & 0.025 & 0.226 & 0.056 & 0.178 & 0.044 \\
251 & 2000251 & 55351.026065 & 86 & 33.80 & 3.4 & 367.85 & 19.5 & 0.887 & 0.027 & 0.867 & 0.014 & 0.878 & 0.057 & 0.904 & 0.034 & 0.163 & 0.050 & 0.129 & 0.039 \\
254 & 2000254 & 55307.007853 & 60 & 12.41 & 0.5 & 384.11 & 5.5 & 0.832 & 0.018 & 0.898 & 0.003 & 0.847 & 0.034 & 0.918 & 0.019 & 0.227 & 0.055 & 0.178 & 0.043 \\
262 & 2000262 & 55325.475207 & 39 & 15.17 & 0.1 & 408.46 & 1.2 & 0.828 & 0.004 & 0.880 & 0.002 & 0.823 & 0.009 & 0.929 & 0.005 & 0.225 & 0.052 & 0.177 & 0.041 \\
272 & 2000272 & 55308.896108 & 34 & 36.53 & 2.5 & 331.60 & 10.2 & 0.953 & 0.008 & 0.915 & 0.012 & 0.887 & 0.036 & 0.913 & 0.017 & 0.071 & 0.019 & 0.056 & 0.015 \\
\enddata
\end{deluxetable}
\end{longrotatetable}

\begin{figure}[htb]
  \begin{center}
    \includegraphics[width=5in]{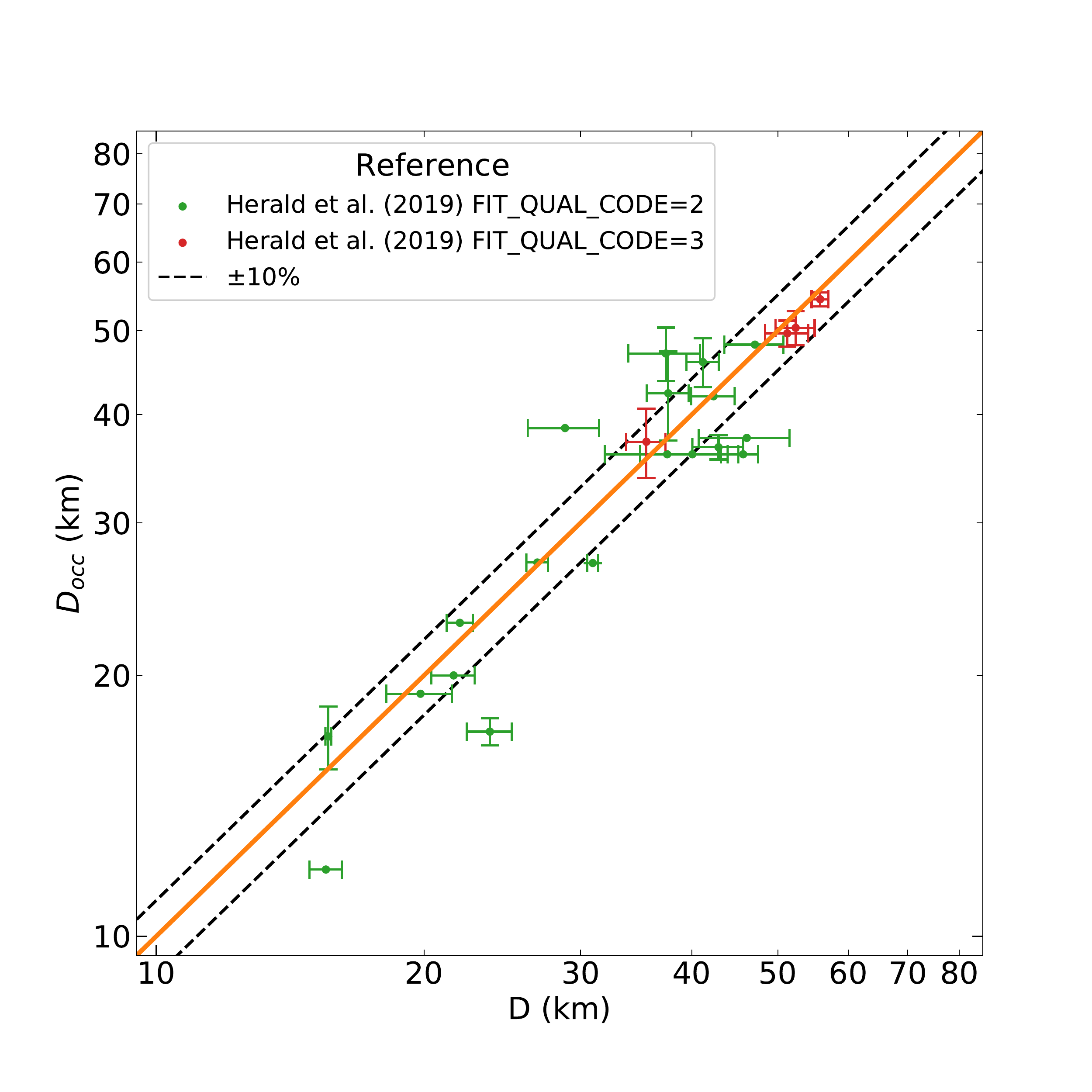}
\caption{Comparison of diameter estimates from stellar occultation observations to the results presented in this work.  \citet{PDSocc} did not report uncertainties for the data points without vertical error bars.  The agreement between the two data sets is quantified by a mean normalized orthogonal distance metric $\langle d_{\perp} \rangle$ of \Dperp (see text).  Asteroid 874 has a stellar occultation diameter of 51.21 $\pm$ 30.9 km \added{with a quality code of 3} and is not plotted \added{for clarity} but is included in the $\langle d_{\perp} \rangle$ calculation.}
\label{fig:occ_compare}
\end{center}
  \end{figure}

To help validate our results, we compared the diameter values obtained
with our model fits to a compilation of occultation diameters
submitted to the Planetary Data System (PDS) \citep{PDSocc}. For this
comparison (Figure \ref{fig:occ_compare}), we used only measurements
with the highest occultation quality code values of 2 or 3.
When two dimensions were reported (ellipse major and minor axes $2a$ and $2b$), we calculated the diameter of a sphere with
the equivalent volume, specifically
$D_{\rm occ} = 2 (a^2 \times b)^{1/3}$, because the
orientation of the body at the time of the WISE observations cannot be
assumed to be identical to the orientation at the time of the
occultation observations.

We quantified the agreement between the two data sets by calculating
a mean normalized orthogonal distance metric $\langle d_{\perp} \rangle$:
\begin{equation}\label{eq:D_perp}
 \langle d_{\perp} \rangle = \frac{1}{N}\sum_{i=1}^{N} \frac{|D_{{\rm occ}, i} - D_i|}{\sqrt{{\sigma_{{\rm occ}, i}}^2 + \sigma_i^2}},
\end{equation}
where the summation is carried out over all $N$ comparisons,  %
$D_{{\rm occ}, i}$ and $\sigma_{{\rm occ}, i}$ are the reported occultation 
diameters and errors, respectively, and $D_i$ and $\sigma_i$ 
are the diameters and errors obtained in this work.
Some occultation-based diameters were reported without
associated uncertainties.  In these situations, we assigned the median
uncertainty associated with the other occultation diameters in the
comparison set ($\sim2$ km).  We found a value of
$\langle d_{\perp} \rangle = \Dperp \ (N=24)$ when comparing our results to the stellar occultation results, which indicates good agreement.

Note that diameter comparisons for large asteroids
are not available to us because WISE observations of large asteroids
are generally saturated in W3, and sometimes also in W4.  In this
study, we focused on the best data and rejected WISE observations
reported with a nonzero saturation flag
as potentially unreliable.  The NEOWISE
team took a different approach.  They applied a linear correction to
saturated W3 observations \citep{main11calibration,masi11}.  The
analytical expression for this correction
was not published until a \replaced{preprint}{withdrawn manuscript} by
\citet{wrig18}, but the method for determining the coefficients in this expression was not described.
In a conference paper,
\citet{wrig19aas}
examined a very similar linear correction term for W3
by requiring that NEOWISE model estimates match known
diameters for about a hundred large asteroids.  However, it is not
stated whether the methodology is the same as that used in the earlier
NEOWISE papers.
If the known
asteroid diameters were used to calibrate the linear correction, then one
obviously
could not use modeling results from the corrected examples to evaluate
NEOWISE modeling accuracy.  However there is a larger problem with the
concept of applying a correction to saturated observations: very few
of the asteroids observed by WISE have saturated W3 observations.  Out
of about 1.4 million W3 observations that meet the filter criteria of
this work (except for the saturation filter), only 658 observations of
132 asteroids are identified by the WISE pipeline as saturated in W3.
So even if a W3 saturation correction were to be valid, it would not
be widely applicable.

We also quantified the agreement between stellar occultation diameters
and the 4-band fit solutions posted by the NEOWISE team to the PDS
\citep{PDSNEOWISE}.  We used the set of 
19
asteroids that are common
to NEOWISE, our solutions, and the solutions of
\citet{PDSocc}.
We found values of $\langle d_{\perp,{\rm this\ work}} \rangle = \DperpNeoThis \ (N=19)$ and
$\langle d_{\perp, {\rm NEOWISE}} \rangle = \DperpNeo \ (N=19)$ when comparing the thermal
fit results to the stellar occultation results.  This metric suggests
that our results more closely match stellar occultation results than
the NEOWISE results, by a factor of approximately 2 in this metric.
Some of this difference is related to differences in diameter uncertainties and the fact that the bootstrap method provides more realistic uncertainties than the formal errors of least-squares fits.

\begin{figure}[htbp]
  \begin{center}
    \includegraphics[width=5in]{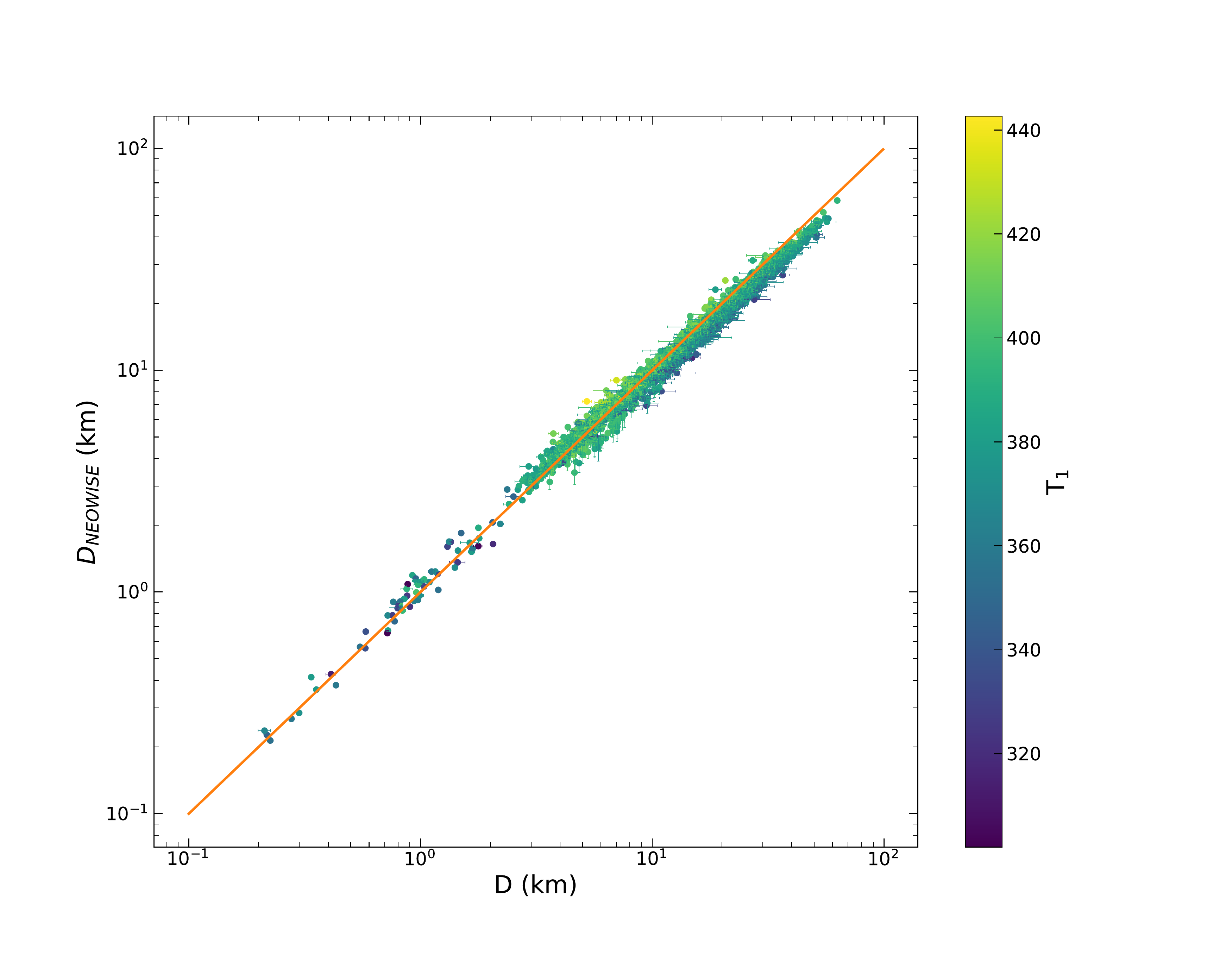}
\caption{Comparison of \nastfitsoverlapNEO NEOWISE 4-band diameter
  results to the results presented in this work.  The best-fit $T_1$
  parameter is shown with a color scale.  The apparent horizontal
  structure is related to the fact that NEOWISE diameter uncertainties
  are generally $\sim$3 times smaller than the more realistic bootstrap uncertainties used in this work.}
\label{fig:NEOWISE_compare}
\end{center}
  \end{figure}

We also compared our diameter results and associated uncertainties
directly to the results posted to the PDS by the NEOWISE team
\citep{PDSNEOWISE}.  A total of \nastfitsoverlapNEO model fits of
\nastsoverlapNEO unique asteroids (2,580 MBAs and 71 NEAs) were common
between the NEOWISE results and our results.\footnote{There are 163
entries in the NEOWISE MBA PDS archive with strings that do not match
any designation known to the MPC, HORIZONS, or the online WISE All-Sky
Single Exposure (L1b) Source Table.  Among these, 84 entries appear to
be incorrect translations of the MPC packed designations and are
recoverable.  The remaining 79 appear to be unrecoverable.} Although
the diameter comparisons are generally favorable
(Figure~\ref{fig:NEOWISE_compare}), there are notable differences.
Some of the diameter discrepancies for some of the asteroids are
likely related to a software bug that, according to \citet{wrig18},
was discovered and fixed in 2011 and affected
more than 100,000 NEOWISE diameter estimates published in 2011.
Inspection of the diameter values affected by the bug indicates that
they have not been corrected in the latest NEOWISE PDS release
\citep{PDSNEOWISE}.
Other sources of discrepancies are related to data selection and model
differences (Sections \ref{sec:data} and \ref{sec:tm}).  We also found
that the NEOWISE estimates of diameter uncertainties are generally 3
times smaller than the bootstrap error estimates.  This finding
suggests that NEOWISE diameter uncertainties may be based on the
incorrect per-observation uncertainties from the WISE pipeline
\citep{hanu15,Myhrvold2018ATM,Myhrvold2018Empirical} as well as on the
formal uncertainties of least-squares fits, which
generally under-estimate actual uncertainties.

\begin{figure}[htbp]
  \begin{center}
    \includegraphics[width=6.5in]{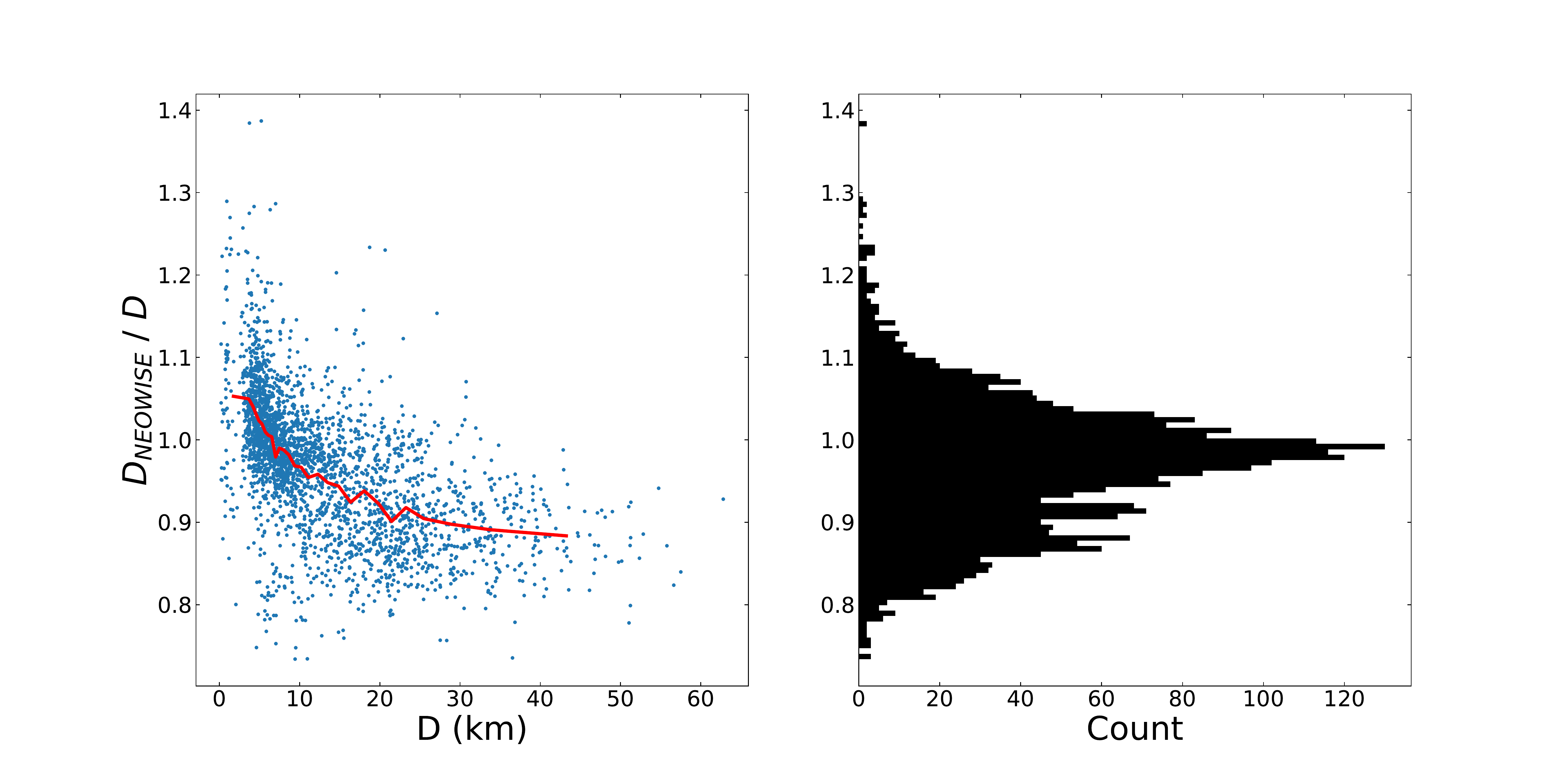}
    \caption{Distribution of NEOWISE 4-band diameters divided by the corresponding diameters obtained in this work.  The red solid line shows the mean ratio obtained in consecutive diameter bins each containing 100 values.
      Approximately 98\%
      of the \nastsoverlapNEO asteroids among our carefully curated sample of \totastcount asteroids have NEOWISE diameter values that agree with ours within $\pm$20\%.
    }
\label{fig:NEOWISE_D_compare}
\end{center}
  \end{figure}

Finally, we computed the ratio of the 4-band NEOWISE diameter values to those
presented in this work.  Despite considerable scatter, a
size-dependent bias
is detectable (Figure \ref{fig:NEOWISE_D_compare}).  The
size-dependent bias could in principle be attributed to the NEOWISE
data analysis, our data analysis, or both.  However, because our
results more closely match the independent size calibration provided by
occultation diameter estimates, we believe that the size-dependent bias is
attributable to the NEOWISE data-processing pipeline, which relies on
the assumption of emissivities of 0.9 in all bands.
The fractions of asteroids whose NEOWISE diameter estimates are in
agreement with our results within $\pm 5\%$, $\pm 10\%$, $\pm 15\%$, and
$\pm 20\%$ are \percentDagreementfive\%, \percentDagreementten\%,
\percentDagreementfifteen\%, and \percentDagreementtwenty\%,
respectively.  The largest differences are $\pm 30\%$.

If we assume that the occultation diameter estimates are exact, we
find that infrared diameter estimates in the comparison sample have a
median error of 9.3\% and a maximum error of 37.7\%.  This comparison
calibrates the oft-stated claim that 4-band NEOWISE diameters are
accurate to 10\% -- the statement appears to be correct in the
best-case scenario of a carefully curated four-band data set with the
techniques described in this work.  However, the statement does not
appear to be correct in general, as evidenced by the fact that only
74\% of NEOWISE four-band diameters agree with ours to $\pm$10\%.  In a
planetary defense context, a 9.3\% error in diameter corresponds to a
28\% error in mass and impact energy.  With a size-dependent bias that
overestimates the diameters of possible impactors, i.e., asteroids
smaller than 5 km, the consequences of the impact hazard calculated
with 4-band NEOWISE diameter estimates may be slightly exaggerated.

\subsection{Albedos}
\begin{figure}[htbp]
  \begin{center}
    \includegraphics[width=6.5in]{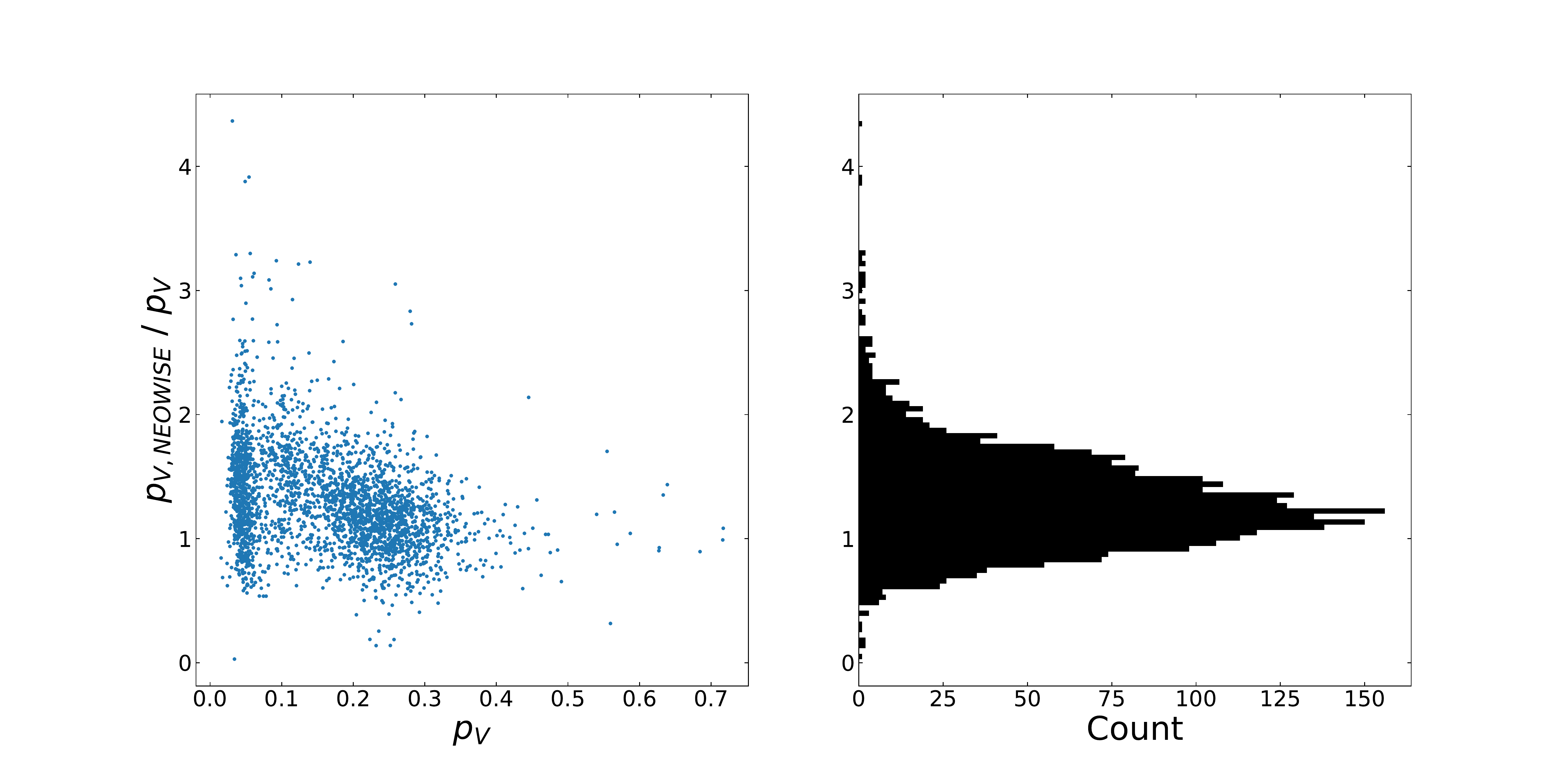}
\caption{Comparison of 2,971 NEOWISE visible albedo 
  results to the results presented in this work.
  In this comparison,
  HORIZONS values of $H$ were used in the calculation of $p_{\rm V}$. 
  Asteroid 2010 MA113\protect\replaced{ with $p_{\rm V, NEOWISE} = -0.999$}{, which does not have a NEOWISE albedo estimate,} is not shown.
}
\label{fig:NEOWISE_pv_compare}
\end{center}
  \end{figure}

\begin{figure}[htbp]
  \begin{center}
    \includegraphics[width=6.5in]{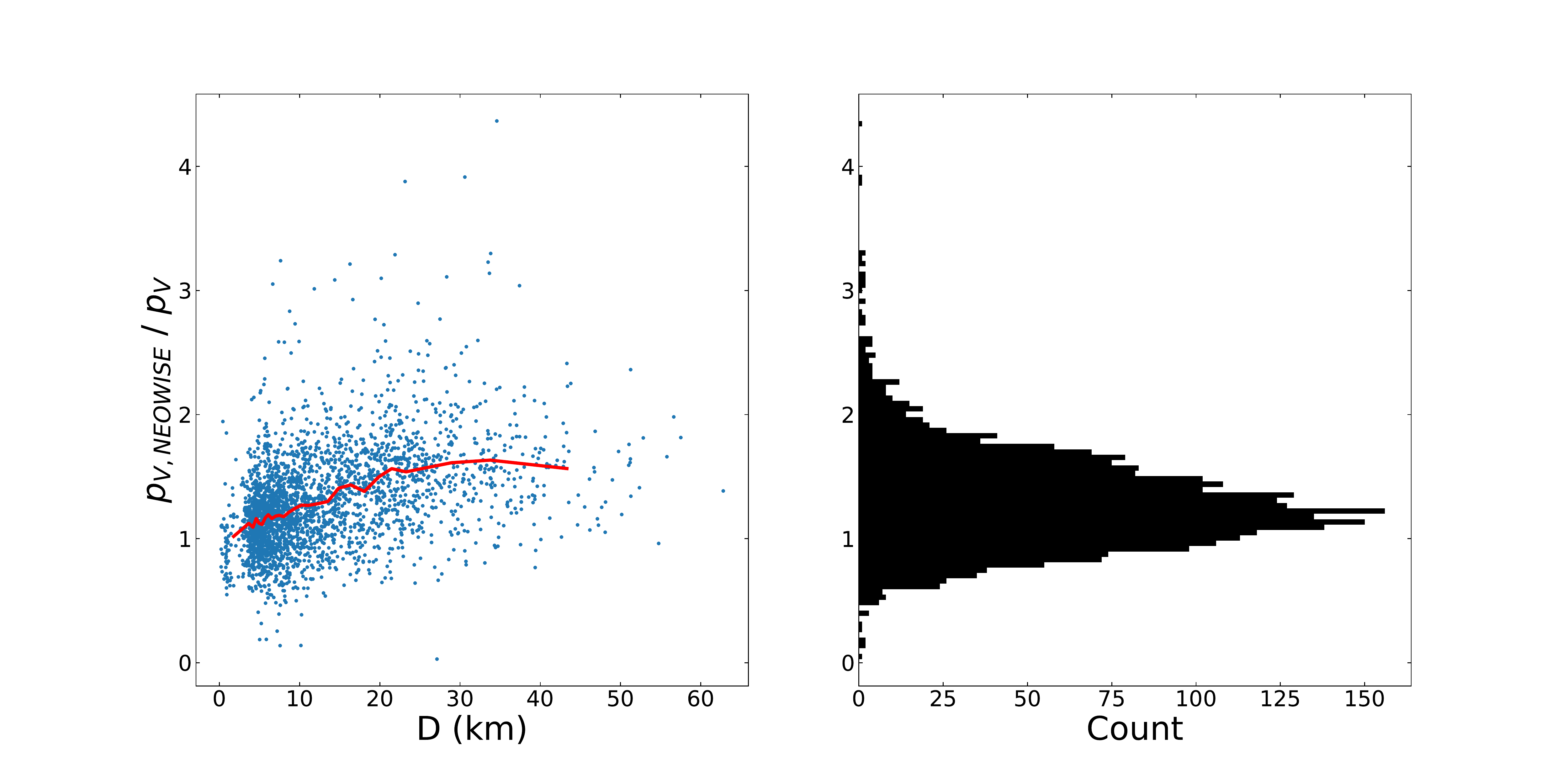}
\caption{Comparison of 2,788 NEOWISE visible albedo 
  results to the results presented in this work, as a function of $D$. The red solid line shows the mean ratio obtained in consecutive diameter bins each containing 100 values.
  In this comparison, \replaced{MPC}{HORIZONS} values of $H$ were used in the calculation of $p_{\rm V}$.
Asteroid 2010 MA113\protect\replaced{ with $p_{\rm V, NEOWISE} = -0.999$}{, which does not have a NEOWISE albedo estimate,} is not shown.
}
\label{fig:NEOWISE_pv_compare_2}
\end{center}
  \end{figure}

We calculated geometric albedo $p_{\rm V}$ values for all asteroids with 
known absolute magnitude $H$ values from
HORIZONS\added{\footnote{JPL's Small Body Database and HORIZONS use the H values computed and published by the Minor Planet Center (MPC) (A. Chamberlin, 2021, personal communication). The MPC improved its algorithm for the calculation of $H$ in the months preceding our data download (F. Spoto, 2021, personal communication). Indeed, the average bias between the MPC $H$ values and the \citet{veres2015} $H$ values for 1314 asteroids in this data set has decreased from 0.30 mag with the legacy algorithm to 0.23 mag with the current algorithm.﻿}}
as well as the definition for visible-band geometric albedo $p_{\rm V}$:
\begin{equation}\label{eq:pv_def}
  p_{\rm V} \equiv \left(\frac{1329}{D}\right)^2 \, 10^{-\frac{2H}{5}},
\end{equation}
where $D$ is the diameter in km \citep{harris2002}.
We performed this calculation with two sets of $H$ values.  First, we
used the $H$ values obtained from
HORIZONS (Table \ref{tbl:results}).
Second, we used the $H$ values reported by \citet{veres2015} if available, otherwise the $H$ values obtained from
HORIZONS
augmented by 0.26
to account for the
mean systematic offset reported by \citet{veres2015}.
Comparisons of
the H$_{\rm HORIZONS}$-based albedo values
to those reported by the NEOWISE team \citep{PDSNEOWISE}
are shown in
Figures \ref{fig:NEOWISE_pv_compare} and
\ref{fig:NEOWISE_pv_compare_2}.

\begin{figure}[htbp]
  \begin{center}
    \includegraphics[width=6.5in]{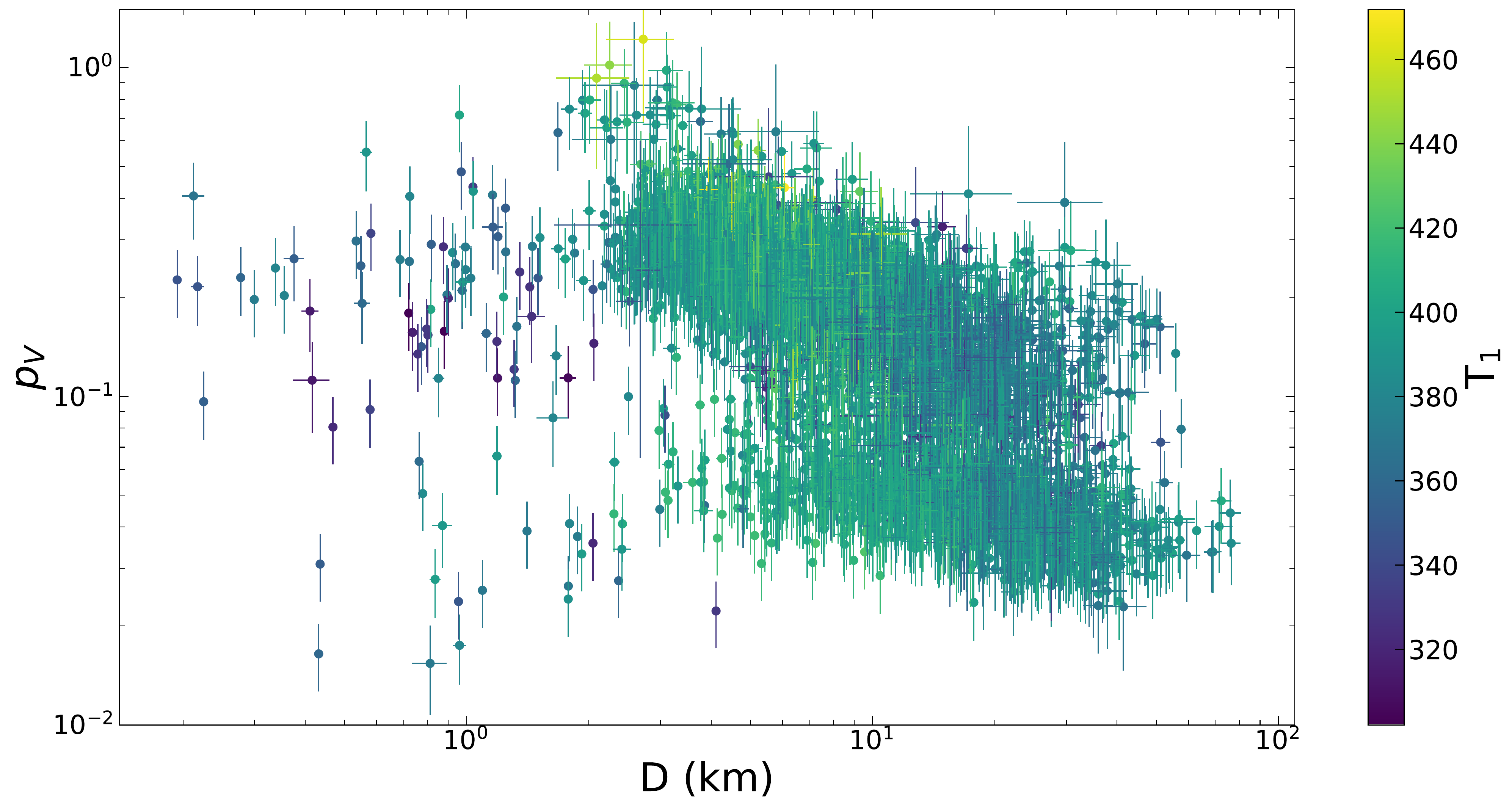}
    \includegraphics[width=6.5in]{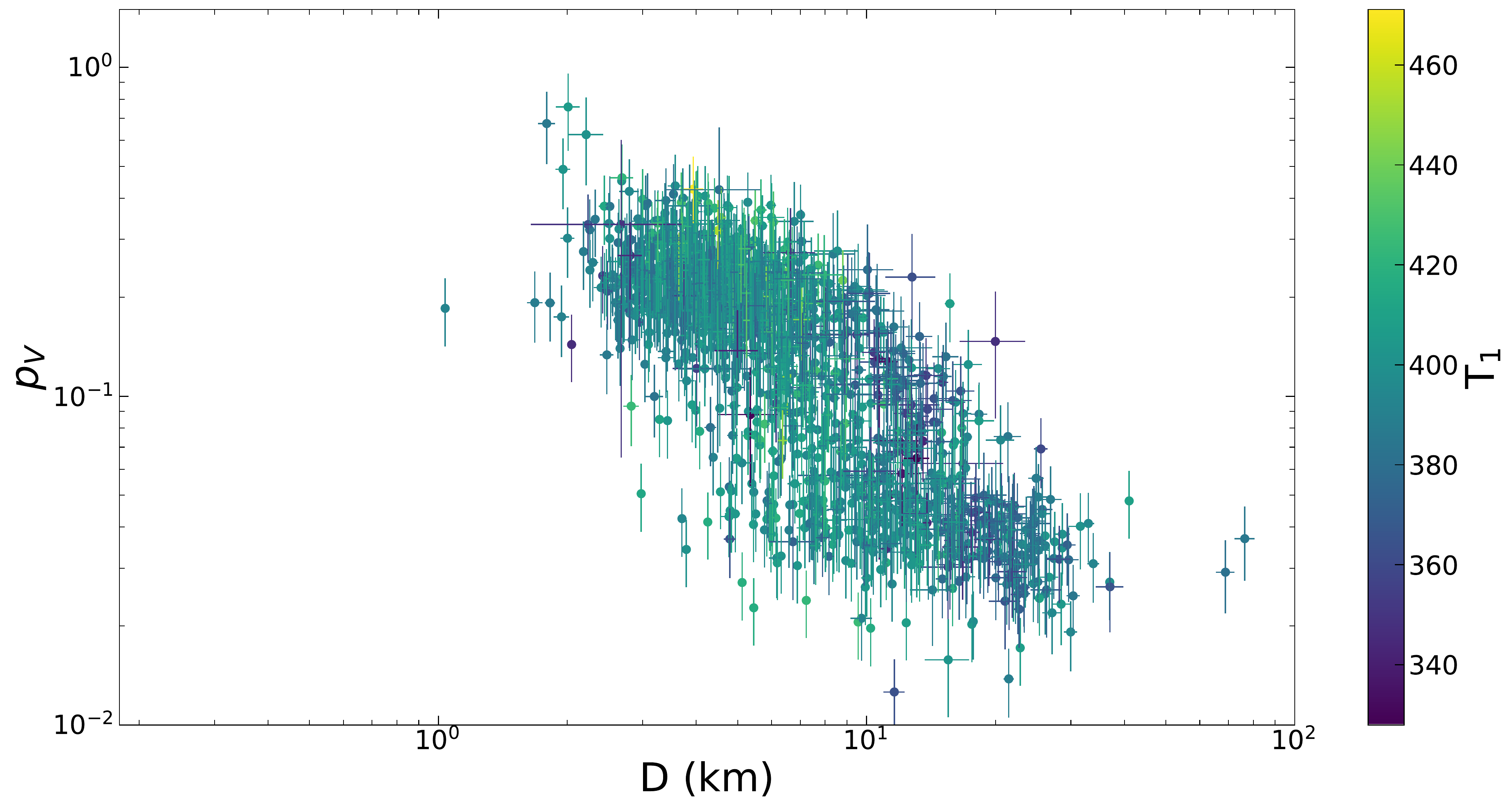}    
\caption{Visible-band geometric albedo $p_{\rm V}$ as a function of the asteroid diameter $D$.  The data are color-coded according to pseudo-temperature $T_1$. We assumed \replaced{an error}{an uncertainty} of 0.25 mag \citep[][Figure 5]{veres2015} on the $H$ values to calculate the error bars on $p_{\rm V}$.  (Top) Albedos calculated with $H$ values obtained from HORIZONS.  (Bottom) Albedos calculated with $H$ values obtained from \citet{veres2015}.}
\label{fig:pv_D}
\end{center}
  \end{figure}

Plots of the visible albedo values as a function of asteroid diameters
(Table \ref{tbl:results}) reveal at least two clusters and an apparent
size-dependence of albedo in each cluster (Figure \ref{fig:pv_D}).
The apparent size dependence persists regardless of the source of the
$H$ values used in the albedo calculation.  However,
\added{size-dependent biases in absolute magnitude estimates have been
  documented \citep{prav12,veres2015} and complicate the
  interpretation of any trend.  Importantly,} the apparent size
dependence is likely due to a selection effect because the optical
surveys and MPCORB database are not complete with respect to the
smaller, darker objects.  Versions of these figures with color-coding
according to phase angle and the quantity $r_{\rm ao} \times r_{\rm
  as}$ are shown in appendix (Figure~\ref{fig:pv_D_cc}).  They reveal
the expected selection effect that objects with the smallest diameters
are detected when observed at the smallest distances, which also tends
to occur at the largest phase angles.  Figure \ref{fig:pv_vs_T1} in
appendix reveals a mild correlation (correlation coefficient 0.2)
between visible-band geometric albedo $p_{\rm V}$ and pseudo-temperature
$T_1$.  Figure \ref{fig:phase_angle_hist} in appendix shows the
distribution of phase angles.

\begin{figure}[htbp]
  \begin{center}
        \includegraphics[width=6.5in]{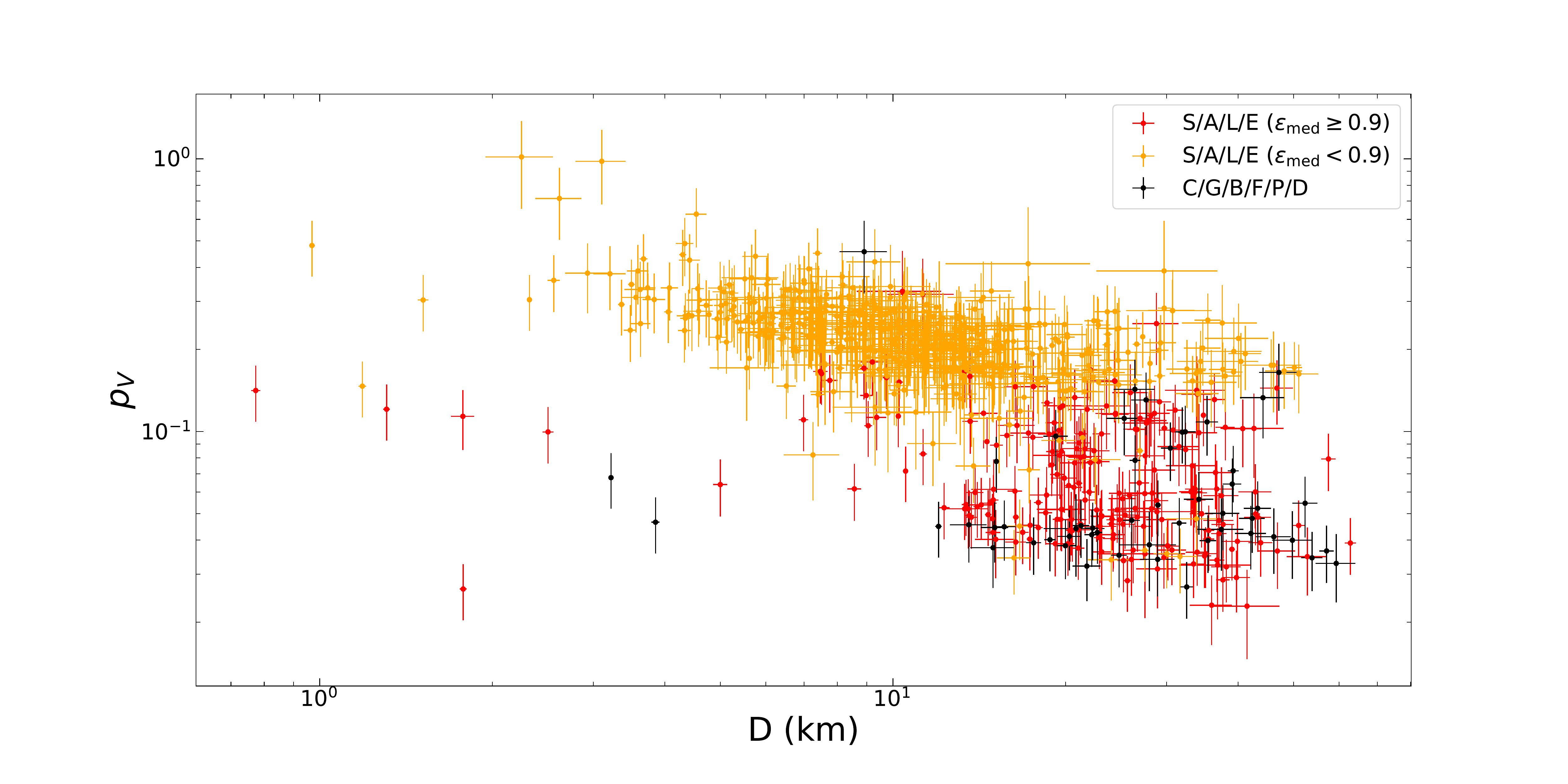}
\caption{Visible-band geometric albedo $p_{\rm V}$ as a function of the asteroid diameter $D$ for \naststax asteroids with known taxonomic classifications.  S/A/L/E-class asteroids are split according to their median infrared emissivity (Section~\ref{sec-eps}).  Albedos were calculated with $H$ values obtained from HORIZONS.}
\label{fig:pv_D_tax}
\end{center}
  \end{figure}

Only about 18\% of asteroids in our sample have known taxonomic
classes according to the latest collection available on the NASA
Planetary Data System\citep{taxo}.  The diameters and albedos of
\naststax asteroids with known taxonomic classes are shown on
Figure~\ref{fig:pv_D_tax}.  As expected, asteroids with classes
traditionally thought to correspond to darker asteroids (C/G/B/F/P/D)
have low albedos.  However, asteroid with classes traditionally
thought to correspond to lighter asteroids (S/A/L/E) have albedos
that span the full range of observed values.  The low-albedo asteroids
in the S/A/L/E classes tend to be associated with high median infrared
emissivities (Section~\ref{sec-eps}).  The ``C'' classification of the
high-albedo asteroid 1653 Castafiore appears to be an outlier.

\begin{figure}[htbp]
\begin{center}
	\includegraphics[width=6.5in]{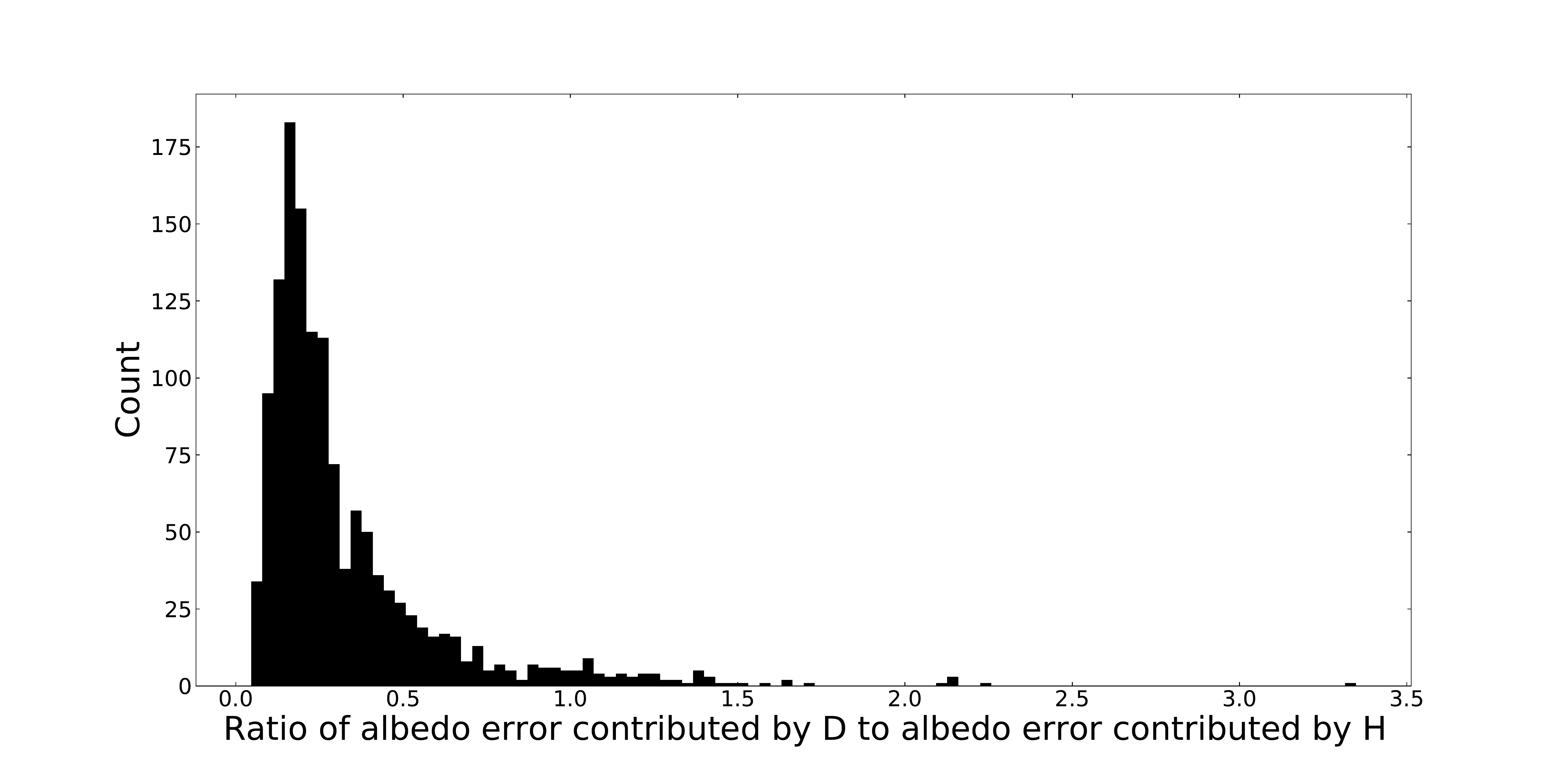}   
\caption{Ratio of the two terms used to calculate the visible-band geometric albedo $p_{\rm V}$ error. We assumed \replaced{an error}{an uncertainty} of 0.25 mag \citep[][Figure 5]{veres2015} on the $H$ values. Errors on the diameter estimates were obtained by calculating the standard deviations of \nbootstraptrials bootstrap resampling solutions.  Note that the error on $H$ is the dominant contribution for $> 91\%$ of the asteroids.
\label{fig:DH_hist}}
\end{center}
\end{figure}

Figure \ref{fig:DH_hist} indicates that the errors on the visible 
albedo values are dominated by the uncertainty on the $H$ measurement, 
which we assumed to be 0.25 mag for all asteroids \citep[][Figure 5]{veres2015}.  \citet{prav12} described 0.2 mag as the smallest plausible uncertainty on absolute magnitudes for survey data and \citet{masi21} argued that these uncertainties will likely reach up to $\sim$1 mag for objects observed at large phase angles.  We found that $H$ uncertainties largely dominate the albedo error budget even with a generous assumption on the quality of $H$ values.

\subsection{Emissivities}
\label{sec-eps}
\begin{figure}[p!]
  \begin{center}
    \includegraphics[width=6.5in]{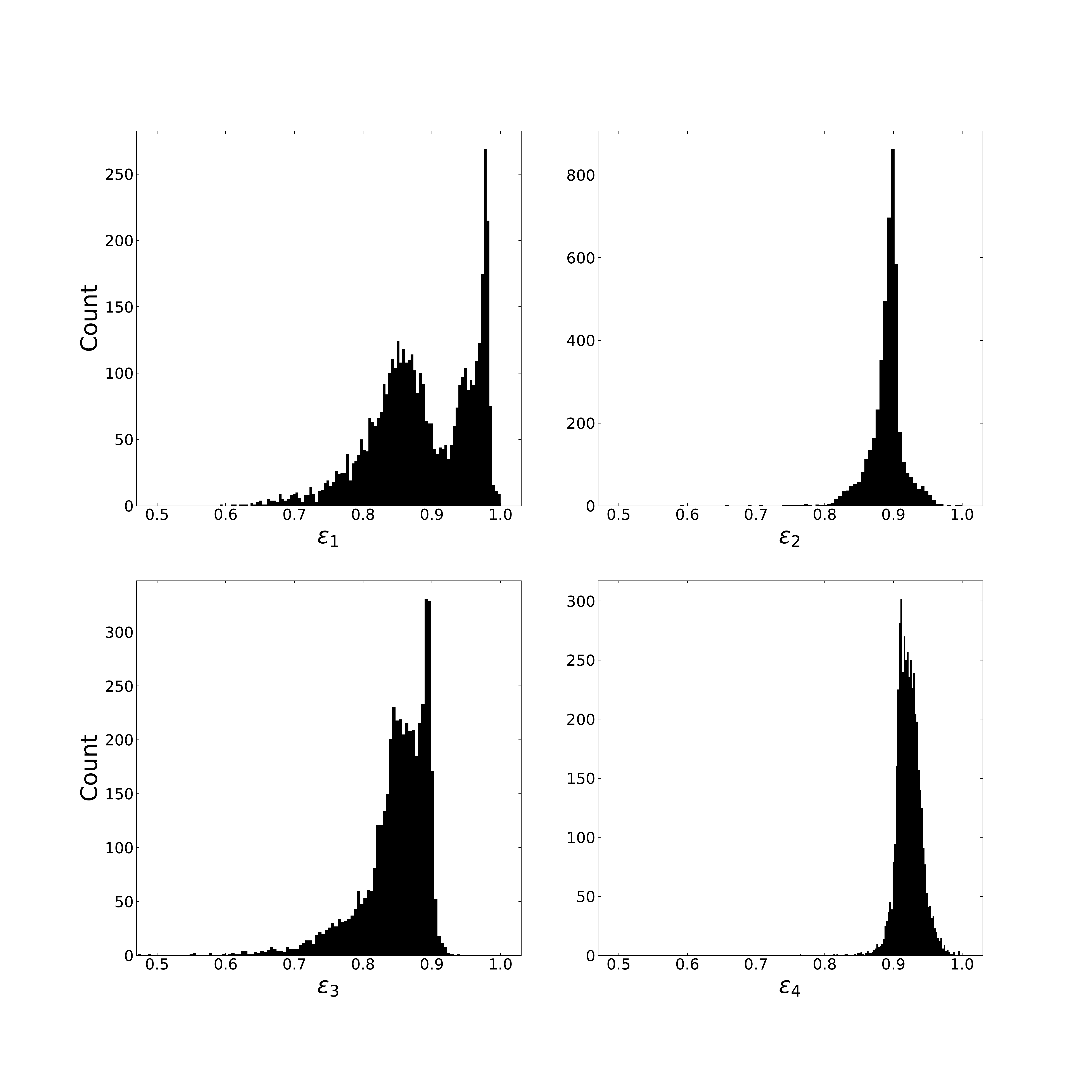}
\caption{Histograms of best-fit emissivity values for each of the WISE bands for \totastcount asteroids.  %
}
\label{fig:eps_hist}
\end{center}
  \end{figure}

The distributions of emissivity values obtained with our asteroid
thermal model fits are shown in Figure \ref{fig:eps_hist}.  Although
our regularized fitting algorithm does favor small dispersions among
emissivities, the correlation matrix (Table~\ref{tab-corr}) indicates
that there was sufficient freedom for individual emissivities to take
on disparate values.  The emissivity in W2 does not correlate strongly
with emissivities in any other band.  Band W2 behaves somewhat
differently from other bands in that W2 flux includes varying
contributions from emitted and reflected light depending on observing
circumstances, whereas W1 flux is almost entirely due to reflected
light and W3--4 fluxes are almost entirely due to thermal
emission.

\begin{table}[h]
  \centering
  \begin{tabular}{c|cccc}
      &      $\epsilon_1$  &       $\epsilon_2$  &       $\epsilon_3$  & $\epsilon_4$ \\
\hline
$\epsilon_1$ &  1.    &  0.241 &   0.758 & -0.157 \\
$\epsilon_2$ &  0.241 &  1.    &   0.171 &  0.152 \\
$\epsilon_3$ &  0.758 &  0.171 &   1.    & -0.526 \\
$\epsilon_4$ & -0.157 &  0.152 &  -0.526 &  1.         \\
  \end{tabular}
  \caption{Correlation matrix of best-fit infrared emissivities for \totastcount asteroids.}
  \label{tab-corr}
  \end{table}

\begin{figure}[h]
  \begin{center}
    \includegraphics[width=3.5in]{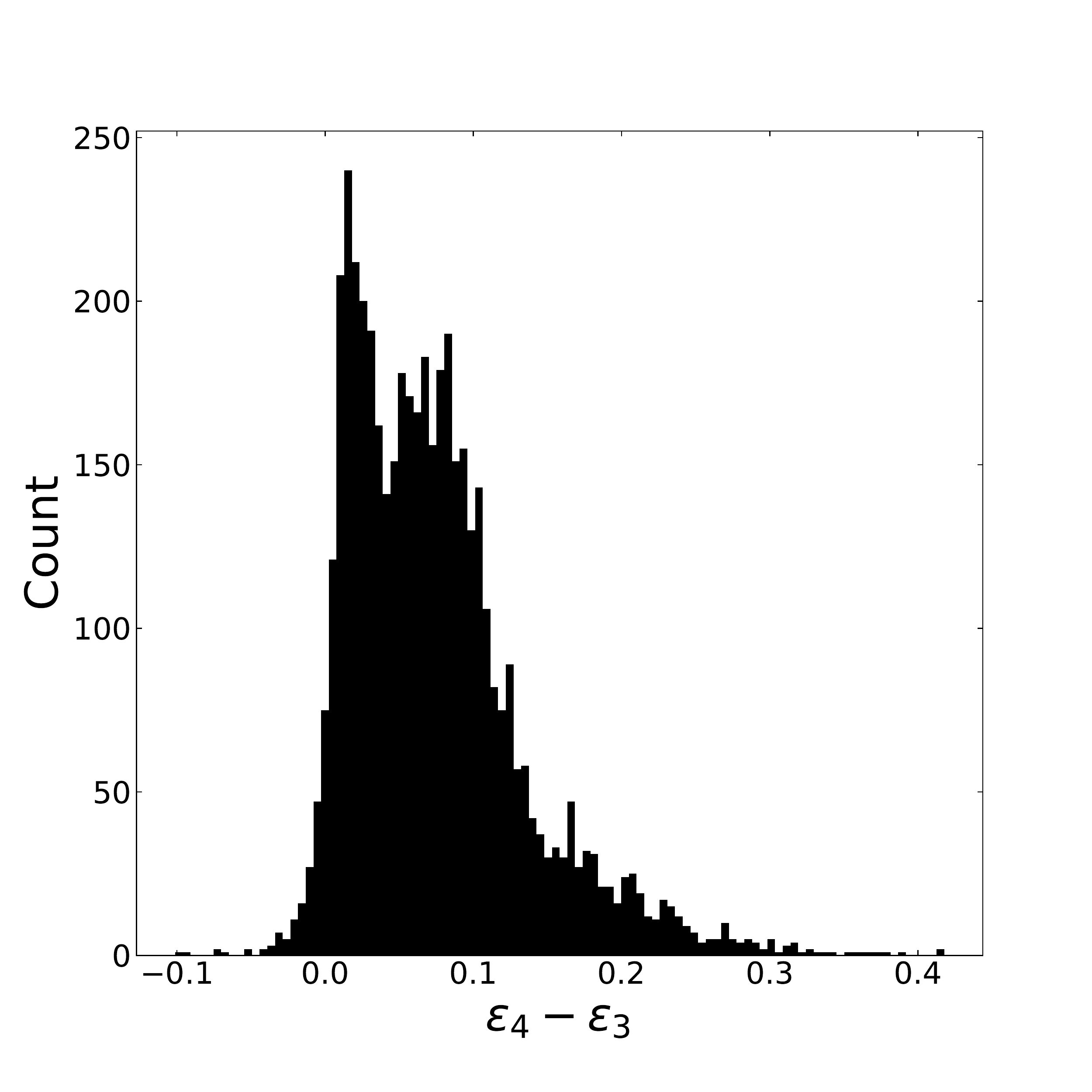}
\caption{Histogram of $\epsilon_4 - \epsilon_3$ values for \totastcount asteroids.}
\label{fig:eps4_minus_eps3}
\end{center}
  \end{figure}

We examined the distribution of $\epsilon_4 - \epsilon_3$ values
(Figure~\ref{fig:eps4_minus_eps3}).  The distribution is asymmetric
and indicates generally higher emissivity values in W4. This
observation reflects differences in physical properties at these two
wavelengths and is consistent with lab spectra of chondritic
materials~\citep[e.g.,][Figure 1]{Myhrvold2018ATM}.

\begin{figure}[p]
  \begin{center}
    \includegraphics[width=6.5in]{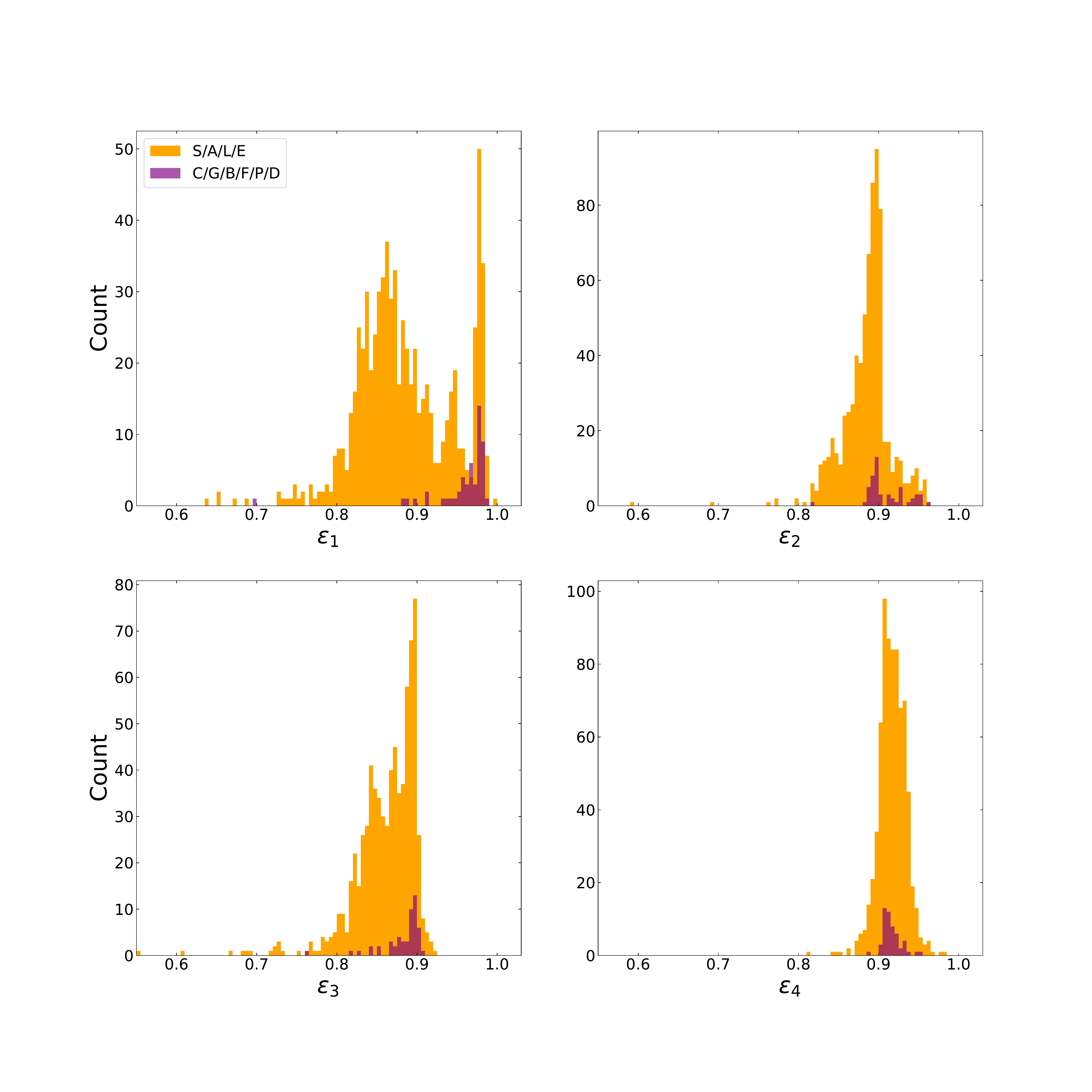}
\caption{Histograms of best-fit emissivity values for each of the WISE bands for \naststax asteroids with known taxonomic type.}
\label{fig:eps_hist_tax}
\end{center}
  \end{figure}

We examined the emissivities of \naststax asteroids with known taxonomic classes (Figure~\ref{fig:eps_hist_tax}).  As expected, asteroids with classes traditionally thought to correspond to darker asteroids (C/G/B/F/P/D) have high emissivity values.  However, asteroid with classes traditionally thought to correspond to lighter asteroids (S/A/L/E) have emissivities that span the full range of observed values.

\begin{figure}[h]
  \begin{center}
    \includegraphics[width=6in]{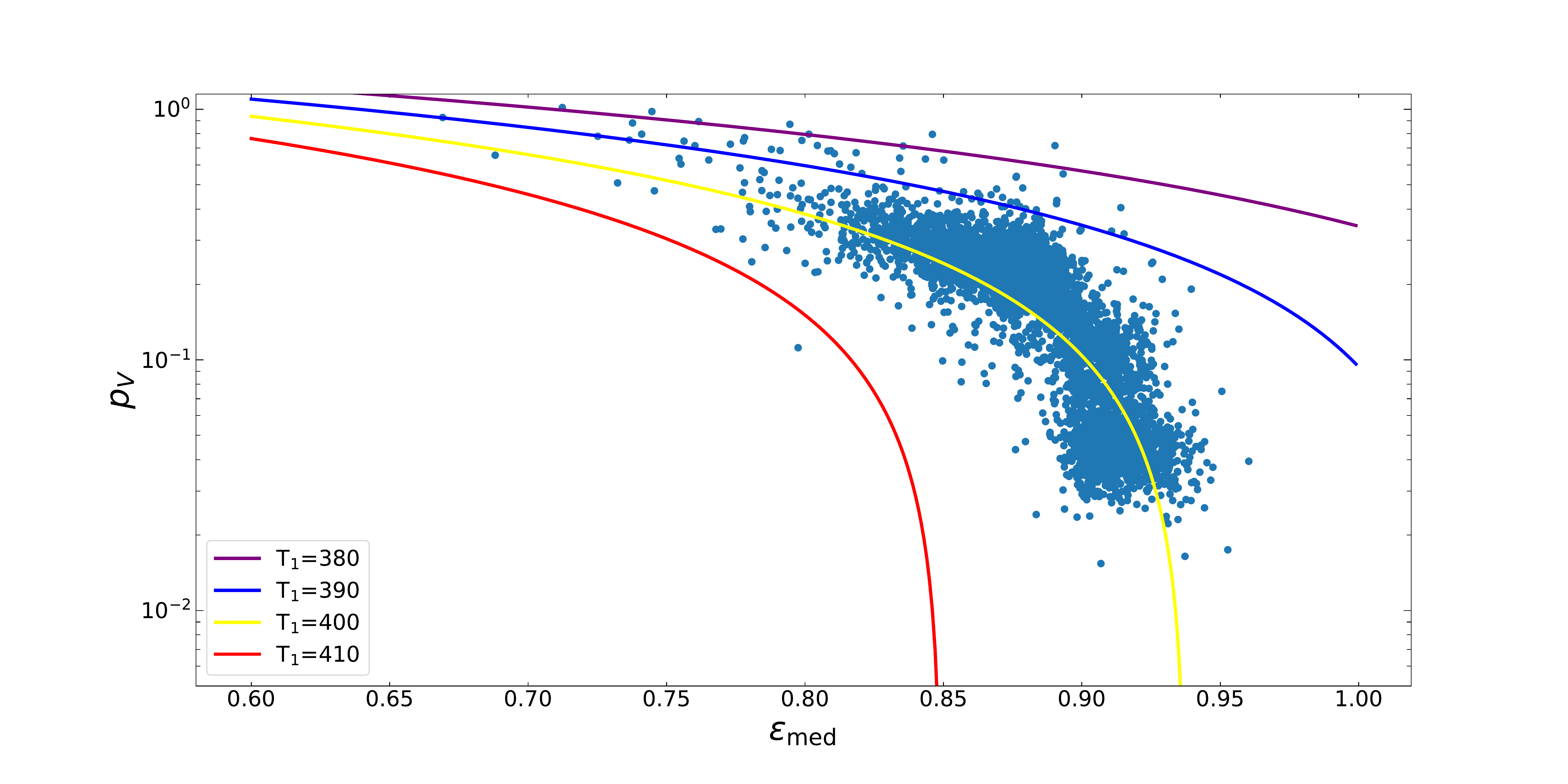}
\caption{Relationship between visible-band geometric albedo $p_{\rm V}$ and median infrared emissivity for \totastcount asteroids. 
  The colored curves represent equation (\ref{eq:T1}) with $q=0.384$, $\eta=1$, and various values of the pseudo-temperature $T_1$.
}
\label{fig:pv_vs_eps}
\end{center}
  \end{figure}

The median infrared emissivity calculated from the best-fit
$\epsilon_i$ $(1 \leq i \leq 4)$ values is highly (negatively)
correlated with the visible-band geometric albedo $p_{\rm V}$, with a
correlation coefficient of -0.842.  The infrared emissivities and the
visual albedo are connected through the assumption that the Bond
albedo $A \approx A_V$ and equation (\ref{eq:T1}), as illustrated on
Figure \ref{fig:pv_vs_eps}.  The emissivity assumptions therefore have
an impact on calculated albedo values.  For instance, the albedo of an
asteroid with best-fit median emissivity of 0.93 would go from
$\sim$2\% to $\sim$10\% if we forced its emissivity to 0.9.
Conversely, promoting high emissivities through a regularization
method may result in lower albedos.  These trends may explain in part
the apparent dichotomy in albedos of S/A/L/E-class asteroids observed
on Figure~\ref{fig:pv_D_tax}.

Figure \ref{fig:eps_med_vs_T1} in appendix reveals a mild negative
correlation (correlation coefficient -0.27) between the median
emissivity value and pseudo-temperature $T_1$.

\subsection{Beaming Parameter}
\label{sec-eta}
The procedure used in this work does not fit for the beaming parameter
$\eta$, which has a variety of obscure physical interpretations
\citep{Myhrvold2018ATM}.  Instead, we absorb it into a
pseudo-temperature whose physical interpretation is straightforward.
For completeness and for comparison to other studies that use this
parameter, we used equation (\ref{eq:T1}) to estimate the value of the
beaming parameter and verify its behavior as a function of
pseudo-temperature $T_1$ (Figure \ref{fig:eta_vs_T1_eps} in Appendix).

\added{\section{Model Limitations}
\label{sec-limitations}

  The NEATM model was built on the assumption of spherical asteroids
  with zero nightside emission and it does not account for thermal
  inertia.  Therefore, the model is likely to underperform for
  irregular or elongated asteroids, for asteroids observed at nonzero
  phase angles, and for asteroids with substantial thermal inertia,
  all of which are represented in the WISE data set.  In particular,
  these errors are likely to be more pronounced for small
  asteroids, because these asteroids have the most elongated and
  irregular shapes; are only detectable at small distances, which
  often results in large phase angles; and appear to have the largest
  thermal inertias \citep{delb15}.  A comprehensive analysis of
  possible errors in NEATM modeling does not appear to be available in
  the literature and is beyond the scope of this work.
  We briefly consider a few aspects.

  Thermal emission measured by WISE may be representative of the
  morning (colder) or afternoon (warmer) side of the asteroid,
  depending on observing circumstances and the spin axis orientation
  of the asteroid.  Higher values of thermal inertia exacerbate the
  morning-to-afternoon asymmetry.  Quantification of this effect
  requires an asteroid thermophysical model \citep{delb15}.

  For a spherical asteroid, the illuminated fraction is $0.5 \,
  (1+\cos{\alpha})$, where $\alpha$ is the Sun-Target-Observer (phase)
  angle.
  The distribution of phase angles for the WISE observations presented
  in this work (Figure \ref{fig:phase_angle_hist}) indicates that the
  vast majority of objects are observed at phase angles $\alpha \leq
  37$ degrees, corresponding to illuminated fractions in excess of
  90\%.  For larger phase angles, the errors due to NEATM's assumption
  of zero nightside emission increase.
  Only 131 (2.8\%) of the observational clusters in our sample were
  obtained at phase angles that exceed 37 degrees.  Among those,
  nearly 90\% correspond to asteroids that are smaller than 3 km, as
  also seen in the color coding of phase angles in Figure
  \ref{fig:pv_D_cc}.  Therefore, we anticipate that phase-angle
  effects mostly affect objects in our sample that are smaller than 3~km.

  WISE detected asteroids with lightcurves of substantial amplitudes
  (Figure~\ref{fig-85839}).  As the asteroid rotates, it presents
  different cross-sectional areas to the observer.  In the simple case
  of an ellipsoid with semi-axes $a$, $b$, $c$ with the $c$ axis in
  the plane of the sky, the apparent cross-section varies between $\pi
  ca$ and $\pi cb$, with a ratio between extrema of $a/b$.  If we
  assume that the reflected and emitted flux are directly proportional
  to the apparent cross section, e.g., if the asteroid surface is
  uniform with a NEATM temperature distribution, then the ratio of
  flux extrema is also $a/b = 10^{0.4A}$, where $A$ is the
  peak-to-peak lightcurve amplitude in magnitude units.  Because NEATM
  does not directly consider lightcurve effects, observations that
  sample the lightcurve inadequately may yield incorrect diameters.
  For instance, the flux observed near the peak of the lightcurve can
  be related to the mean flux with
  \begin{equation}
  \frac{F_{\rm max}}{F_{\rm mean}} \propto \frac{\pi ca}{\pi c(a+b)/2} \simeq 10^{0.2A},
  \end{equation}
  and the diameter inferred from peak flux observations, $D_{\rm max}
  = 2 \sqrt{ca}$, is in error compared to the mean diameter, $D_{\rm
    mean} = 2 \sqrt{c(a+b)/2}$, by a factor of $10^{0.1A}$.  We ran
  Monte Carlo simulations to evaluate diameter errors due to
  incomplete sampling of asteroid lightcurves.  In this idealized
  analysis, we assumed that all WISE bands have the same light curve
  amplitude, and we estimated the diameter from 10,000 averages of
  $N=$ 3, 12, or 46 flux samples randomly selected within a rotation
  period, for a variety of lightcurve amplitudes.  Results are shown
  in table \ref{tab-lc}.  Even for the largest lightcurve amplitude
  $A_{\rm pp}=1.5$ mag, which captures 99.8\% of the asteroids reported in the
  asteroid lightcurve data base~\citep{lcdb}, fractional errors on
  diameters do not exceed 4\% with the minimum number of points ($N$ =
  12) used in this work.  However, errors increase rapidly when fewer
  points are available.  We expect actual errors to be somewhat larger
  because different bands may reveal distinct lightcurves, WISE
  observations may not cover a complete rotation of the asteroid,
  lightcurves can be much more complicated than a simple sinusoid, and
  other simulation assumptions may be violated.
  
\begin{figure}[h]
  \begin{center}
    \begin{tabular}{cc}
    \includegraphics[width=3in]{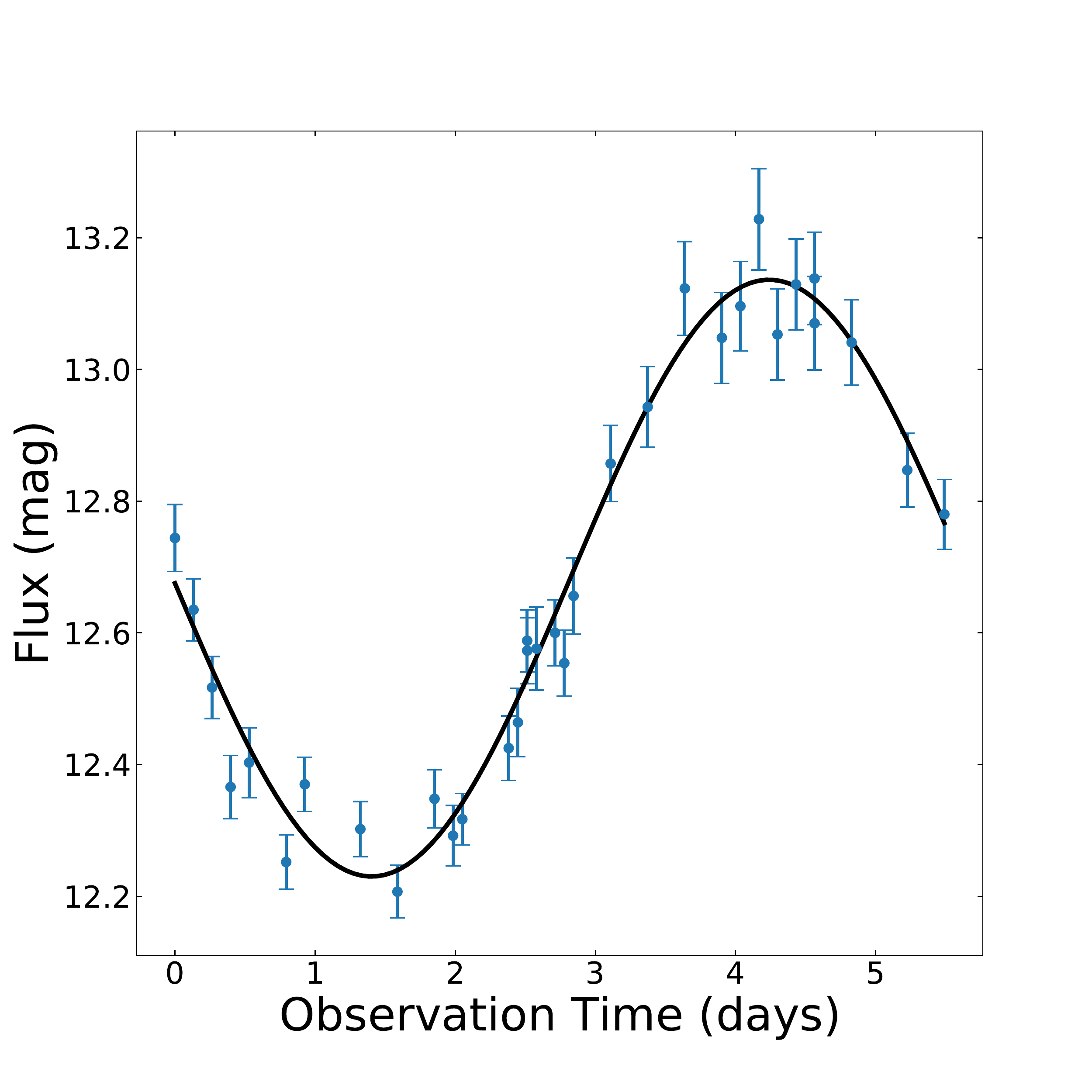} &
    \includegraphics[width=3in]{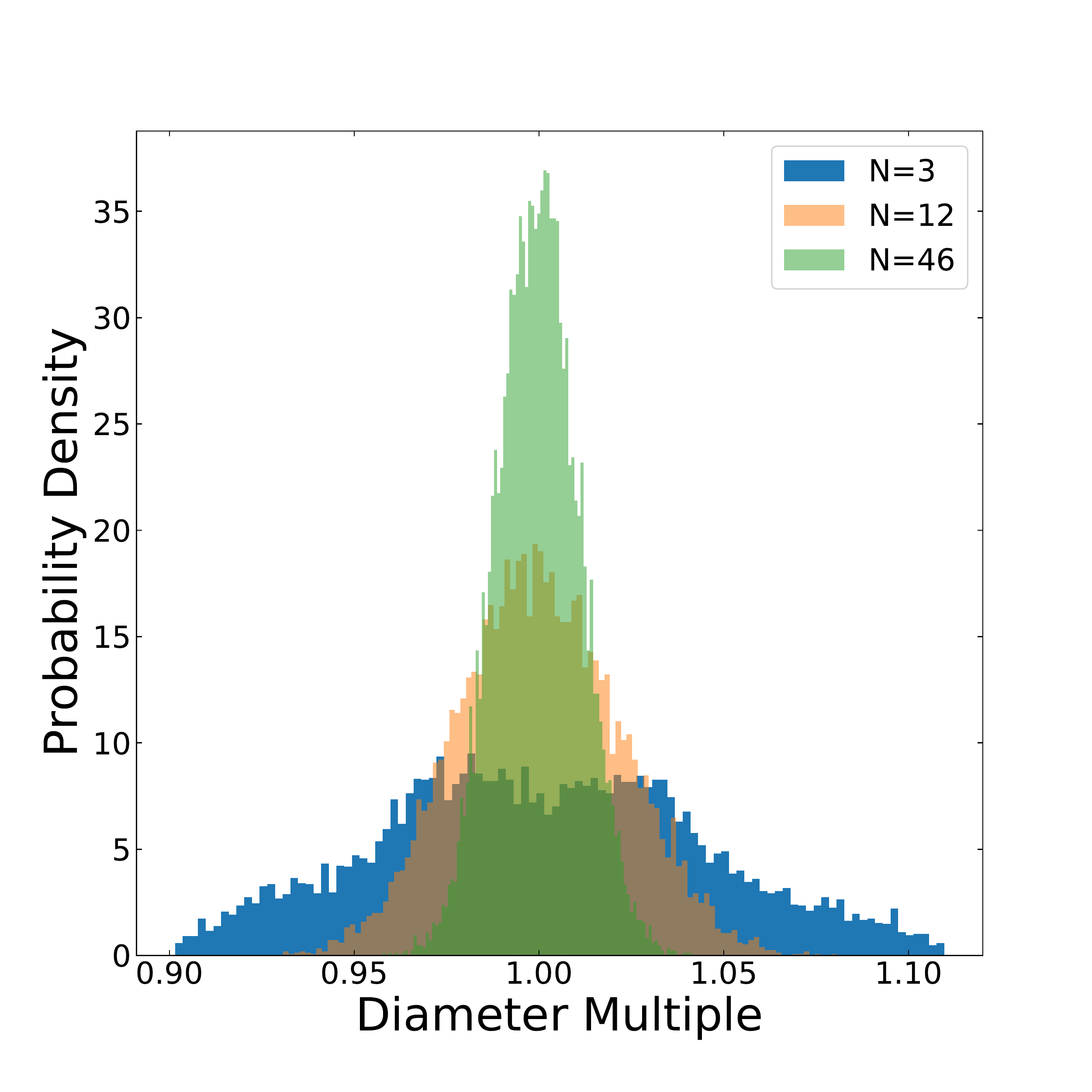}\\
    \end{tabular}
    \caption{(Left) A subset of the W2 WISE observations of asteroid 85839 are well fit by a sinusoidal curve with a peak-to-peak lightcurve amplitude $A_{\rm pp}=0.91$ and a mean magnitude $M=12.7$.
      (Right) Histograms of diameter estimates for 10,000 Monte Carlo realizations obtained with $N=$ 3, 12, or 46 randomly selected flux observations with $A_{\rm pp}=0.91$, expressed as a function of the ``true'' (mean) effective diameter.   We used these simulations to quantify the
      approximate 1-$\sigma$ diameter errors caused by incomplete sampling of asteroid lightcurves.
}
\label{fig-85839}
\end{center}
\end{figure}

\begin{deluxetable}{rrrrrrrrrrrrrrrr}[h!]
  \caption{Fractional 1-$\sigma$ errors (in percent) on effective diameters of elongated objects as a function of peak-to-peak lightcurve amplitude $A_{\rm pp}$ and number of observations $N$. The minimum and median number of observations per cluster in this work are 12 and 46, respectively.  Elongations $a/b$  are also shown.%
  }
  \label{tab-lc}
\tablehead{
\colhead{$A_{\rm pp}$} & \colhead{0.1} & \colhead{0.2} & \colhead{0.3} & \colhead{0.4} & \colhead{0.5} & \colhead{0.6} & \colhead{0.7} & \colhead{0.8} & \colhead{0.9} & \colhead{1.0} & \colhead{1.1} & \colhead{1.2} & \colhead{1.3} & \colhead{1.4} & \colhead{1.5}}
\startdata
$a/b$ & 1.10 & 1.20 & 1.32 & 1.45 & 1.58 & 1.74 & 1.91 & 2.09 & 2.29 & 2.51 & 2.75 & 3.02 & 3.31 & 3.63 & 3.98 \\
\hline
N=3 & 0.48 & 0.96 & 1.43 & 1.91 & 2.41 & 2.86 & 3.29 & 3.83 & 4.33 & 4.76 & 5.24 & 5.81 & 6.13 & 6.71 & 7.12 \\
N=12 & 0.24 & 0.48 & 0.72 & 0.96 & 1.20 & 1.44 & 1.67 & 1.92 & 2.15 & 2.39 & 2.65 & 2.87 & 3.09 & 3.33 & 3.55 \\
N=46 & 0.12 & 0.24 & 0.37 & 0.49 & 0.61 & 0.73 & 0.86 & 0.98 & 1.11 & 1.23 & 1.34 & 1.46 & 1.60 & 1.70 & 1.84 \\
\enddata
\end{deluxetable}

}

\section{Conclusions}\label{sec:conclusions}
We described the results of our analysis of \totobscount WISE
observations of \totastcount asteroids.  In particular, we provided
solutions to 131 asteroids not analyzed by the NEOWISE team and to
1,778 asteroids not analyzed with 4-band data by the NEOWISE team.  The subset
of asteroids was carefully curated to include at least three
observations in each one of the four WISE bands and to eliminate
measurements with artifacts, low S/N, poor photometric quality,
saturation, or questionable PSF fits.  The last two of these filters
remove observations that are likely problematic but were not discarded
in previous work.  We also eliminated sources that were reported
at large distances from the expected ephemeris positions of asteroids
or that experienced conjunction or near-conjunction conditions or
possible background confusion.  These distance-based data filters are
also novel and eliminate incorrect data that were not previously
excluded according to NEOWISE descriptions.  The curated set of
observations provides the best possible scenario for analysis of
four-band infrared data and yields useful performance benchmarks for
diameter, emissivity, and albedo determinations.  Observations in less
favorable conditions, e.g., two-band data, are expected to yield
estimates with larger errors.

We compared best-fit diameter results to independent, high-quality
stellar occultation size estimates.  This calibration revealed that
four-band diameter estimates can be recovered with a median precision
of 9.3\% and a worst-case precision of 37.7\% in this sample, which
correspond to median and worst-case precisions of 28\% and 113\% on
mass and impact energy.

Our results indicate that \added{asteroid diameter estimates can be
  obtained with thermal infrared observations without resorting to
  estimates of the absolute visual magnitude $H$, which are known to
  suffer from a variety of biases, including size-dependent biases.
  In addition, we found that} forcing emissivities to an arbitrary
value of 0.9 can have \replaced{a substantial effect on estimates of
  the visible-band geometric albedo}{detrimental effects on diameter
  and albedo estimates}.  We found that a regularized fitting
algorithm with variable emissivities yields markedly better fits to
the data than a least-squares
fitting algorithm with emissivities arbitrarily set to 0.9.  In
particular, the regularized approach yields diameter estimates that
are $\sim$2 times better than NEOWISE estimates, as quantified by an
orthogonal distance metric with respect to high-quality stellar
occultation diameter estimates.  This result and others are robust
with respect to four different forms of the regularization loss.  The
regularized approach also solved the problem of thermal fits that
completely miss the data in at least one band in a substantial
fraction ($\sim$36\%) of asteroids.
\citet{moey20} reached a similar conclusion by fitting thermal models
that allowed for different emissivities within a Bayesian framework.
Specifically, for a slightly different set of WISE data with different
input filters, \citet{moey20} also
concluded that the
data supported emissivity values other than 0.9 for W3 and W4.

Our results suggest that NEOWISE estimates of diameters and albedos
are affected by size-dependent biases that may pollute estimates of
asteroid size distributions and slightly inflate impact hazard risk
calculations.

We found that emissivities in the
W3 and W4 thermal bands are imperfectly correlated ($\sim$0.53), with a
marked asymmetry revealing generally higher values in W4, consistent
with laboratory spectra for chondritic materials~\citep[e.g.,][Figure
  1]{Myhrvold2018ATM}.

We quantified the sources of error in \replaced{infrared-based estimates of
albedos}{albedo estimates} and found that the primary source of error lies in the
determination of the absolute visual magnitude for over 90\% of
asteroids in this sample.  Because albedo estimates could be readily
improved with better estimates of absolute visual magnitudes, efforts
similar to those of \citet{prav12} and \citet{veres2015} are quite
valuable.

\added{The NEATM model used in this and other works is likely to
  underperform for irregular or elongated asteroids, for asteroids
  observed at nonzero phase angles, and for asteroids with
  substantial thermal inertia.  These errors tend to conspire for
  smaller asteroids, whose infrared-based diameter estimates may be
  less reliable.  We found that phase-angle effects in our curated
  WISE data set are most likely to affect objects with diameters under
  3~km.  Incomplete sampling of asteroid lightcurve variations results
  in additional errors on diameter estimates.  We estimated additional
  fractional errors of 0.24--3.6\% in idealized numerical simulations
  of ellipsoidal asteroids with lightcurves of 0.1--1.5 mag
  amplitudes, respectively, and 12 randomly selected flux estimates.
  Real-world errors due to lightcurve variations are likely somewhat
  larger.}

\acknowledgments We gratefully acknowledge constructive comments from
two reviewers.
We thank Victor Ali Lagoa, Marco Delbo, Alan Harris, Ellen Howell, ﻿Zeljko Ivezic, Joseph Masiero, Joachim Moeyens, Federica Spoto, Peter Veres, and Edward (Ned) Wright for useful discussions.
PP was funded in part by the Nathan P. Myhrvold
Graduate Fellowship.  JLM was funded in part by NASA grant
80NSSC18K0850.

\pagebreak
\appendix
\section{Alternate NEATM geometry} \label{app:neatm}

In some NEATM formulations, a left-handed coordinate system is used
where $\vec{y}$ points at the Sun.  Integration appears to be carried
out over all latitudes and illuminated longitudes, although the
portion of the hemisphere that is not visible to the observer is
removed.  In this formulation, it appears that the observer is
confined to the $yz$ plane if $\theta$ is defined as a latitude and
the $xz$ plane if $\theta$ is defined as a longitude.  We can
reproduce this setup with the following substitutions: $\theta
\leftarrow \phi$ and $\phi \leftarrow 90^{\circ} - \theta$.
The resulting expression is
\begin{equation}\label{eq:NEATM_flux}
  F_{\rm NEATM}(\alpha, \lambda) = \frac{\epsilon(\lambda) D^2}{4 {r_{\rm ao}}^2}  \int_{\phi=-\pi/2}^{\phi=\pi / 2} \int_{\theta=-\pi / 2}^{\theta=\pi / 2} B(\lambda, T(\theta, \phi)) \cos^2 \phi \cos(\theta - \alpha) d\theta d\phi,
\end{equation}
where the portion of the hemisphere not visible to the observer is given zero flux with 
\begin{equation}\label{eq:T_dist}
\begin{cases}
    T(\theta, \phi) = T_{\rm ss}(\cos\phi\cos\theta)^{1/4},& \text{if } \alpha - \pi / 2 \leq \theta \leq \alpha + \pi / 2\\
    0,              & \text{otherwise}
\end{cases}
\end{equation}
Both formulations yield identical thermal flux values for a given set of
input parameters.
\vfill

\section{Alternate forms of the regularization loss} \label{app:alternate}
In the course of our investigation, we attempted a few forms of the
regularization loss (Equation~\ref{eq:R_model}).  In some
formulations, we targeted a norm for the emissivity vector that
corresponds to $\epsilon \sim 0.9$ or $\epsilon \sim 1$ while
minimizing the dispersion between emissivity values with a second term
$\sigma_{\epsilon}$, which is the standard deviation among the
emissivity values:
\begin{equation}\label{eq:R_model1}
  R_{\rm model} = \frac{2 - ||\epsilon||}{2} + 2\sigma_{\epsilon}.
\end{equation}

\begin{equation}\label{eq:R_model2}
  R_{\rm model} = \frac{1.8 - ||\epsilon||}{1.8} + 2\sigma_{\epsilon}.
\end{equation}
\begin{equation}\label{eq:R_model2}
R_{\rm model} = ||\epsilon-0.9|| + 2\sigma_{\epsilon}
\end{equation}

Although we observed slight differences in the diameter and $T_1$
estimates for various versions of the regularization loss, all forms
confirmed our general conclusions.  In particular, all forms exhibited
a bimodal distribution of $T_{\rm 1,min}/T_{\rm 1,max}$ (Figure
\ref{fig:T1_ratios}) and yielded better agreement with occultation
diameters than NEOWISE solutions (Figure \ref{fig:occ_compare}).  All
forms of the loss functions revealed an apparent size-dependent bias
in NEOWISE diameter estimates (Figure \ref{fig:NEOWISE_D_compare}).
All forms also exhibited similar distributions of visible geometric
albedo vs.\ diameter and a taxonomic dichotomy (Figures \ref{fig:pv_D}
and \ref{fig:pv_D_tax}).  Details of the emissivity histograms varied,
but the overall ranges of emissivity values were similar (Figure
\ref{fig:eps_hist}).

We conclude that the details of the regularization loss function have
relatively little impact and that our overall conclusions are robust.  
\vfill

\section{Additional Figures}

\begin{figure}[htbp]
  \begin{center}
    \includegraphics[width=6.5in]{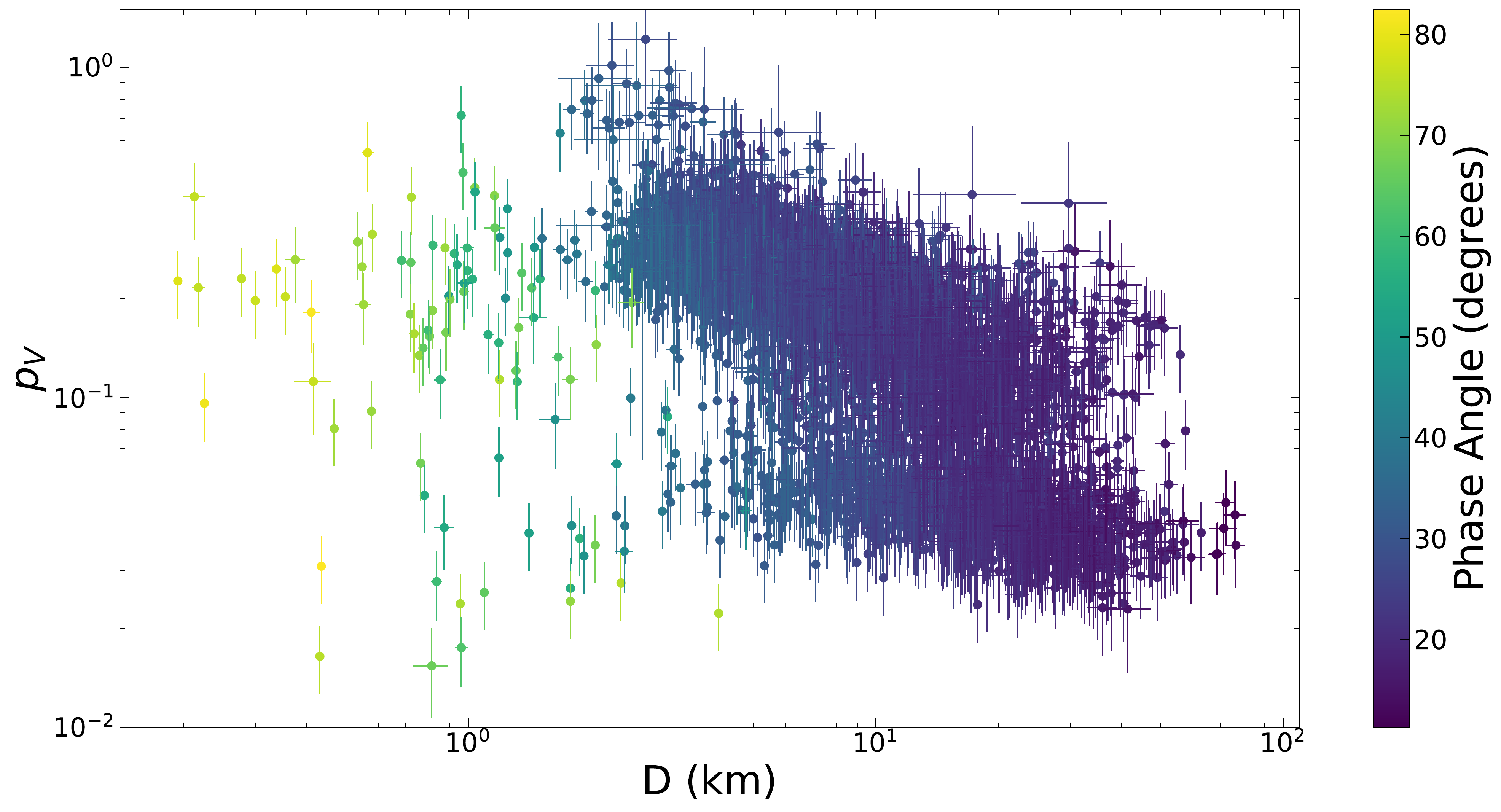}
    \includegraphics[width=6.5in]{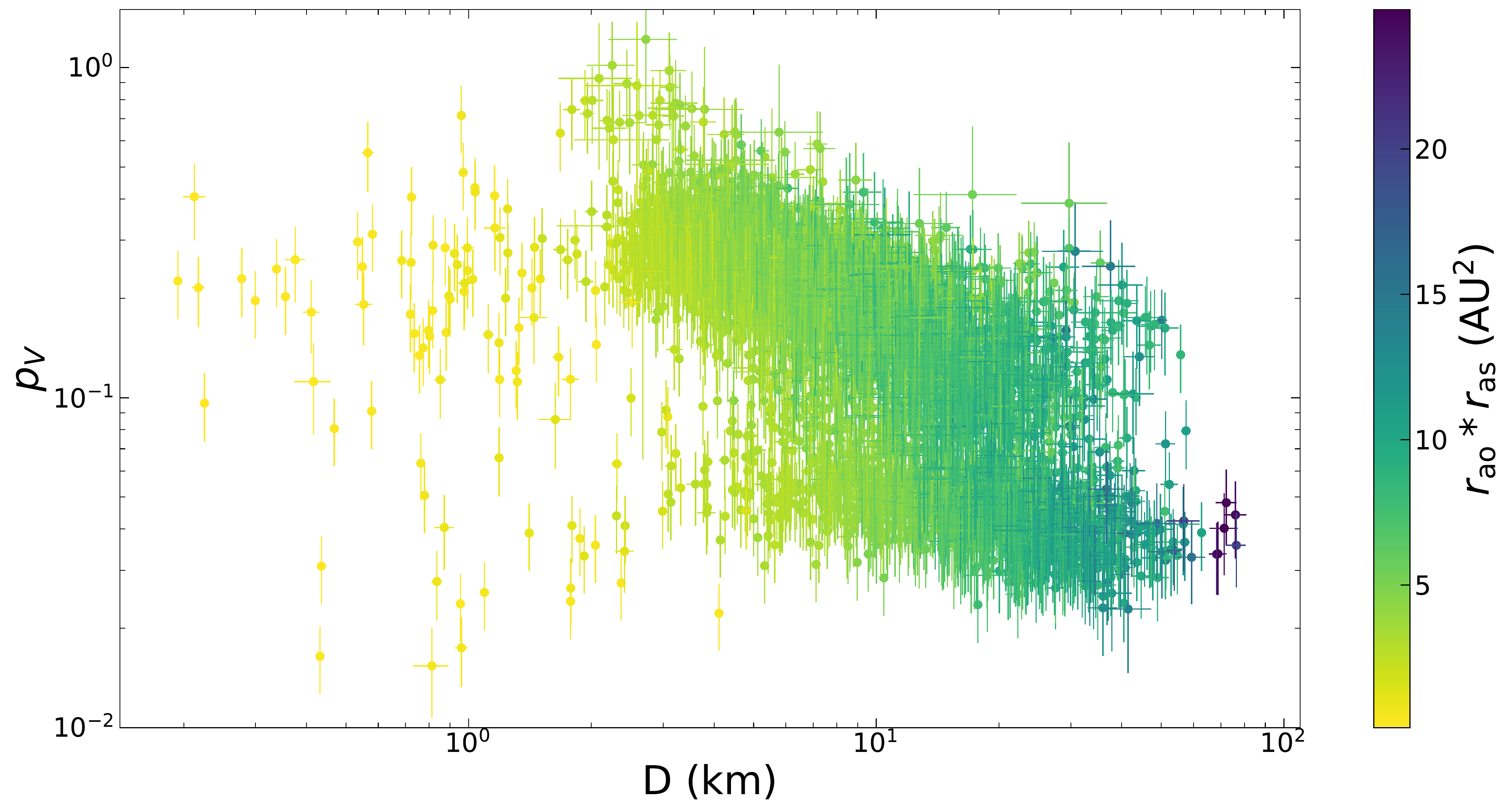}    
\caption{Visible-band geometric albedo $p_{\rm V}$ as a function of the asteroid diameter $D$, color-coded according to the median phase angle (top) and the quantity $r_{\rm ao} \times r_{\rm as}$ (bottom). We assumed \replaced{an error}{an uncertainty} of 0.25 mag \citep[][Figure 5]{veres2015} on the $H$ values to calculate the error bars on $p_{\rm V}$. Albedos are calculated with $H$ values obtained from HORIZONS. \citet{veres2015}.}
\label{fig:pv_D_cc}
\end{center}
  \end{figure}

Figure \ref{fig:pv_D_cc} replicates the top plot of Figure \ref{fig:pv_D}
color-coded according to the median phase angle as well as the quantity 
$r_{\rm ao} \times r_{\rm as}$. 

\begin{figure}[h]
  \begin{center}
    \includegraphics[width=6.5in]{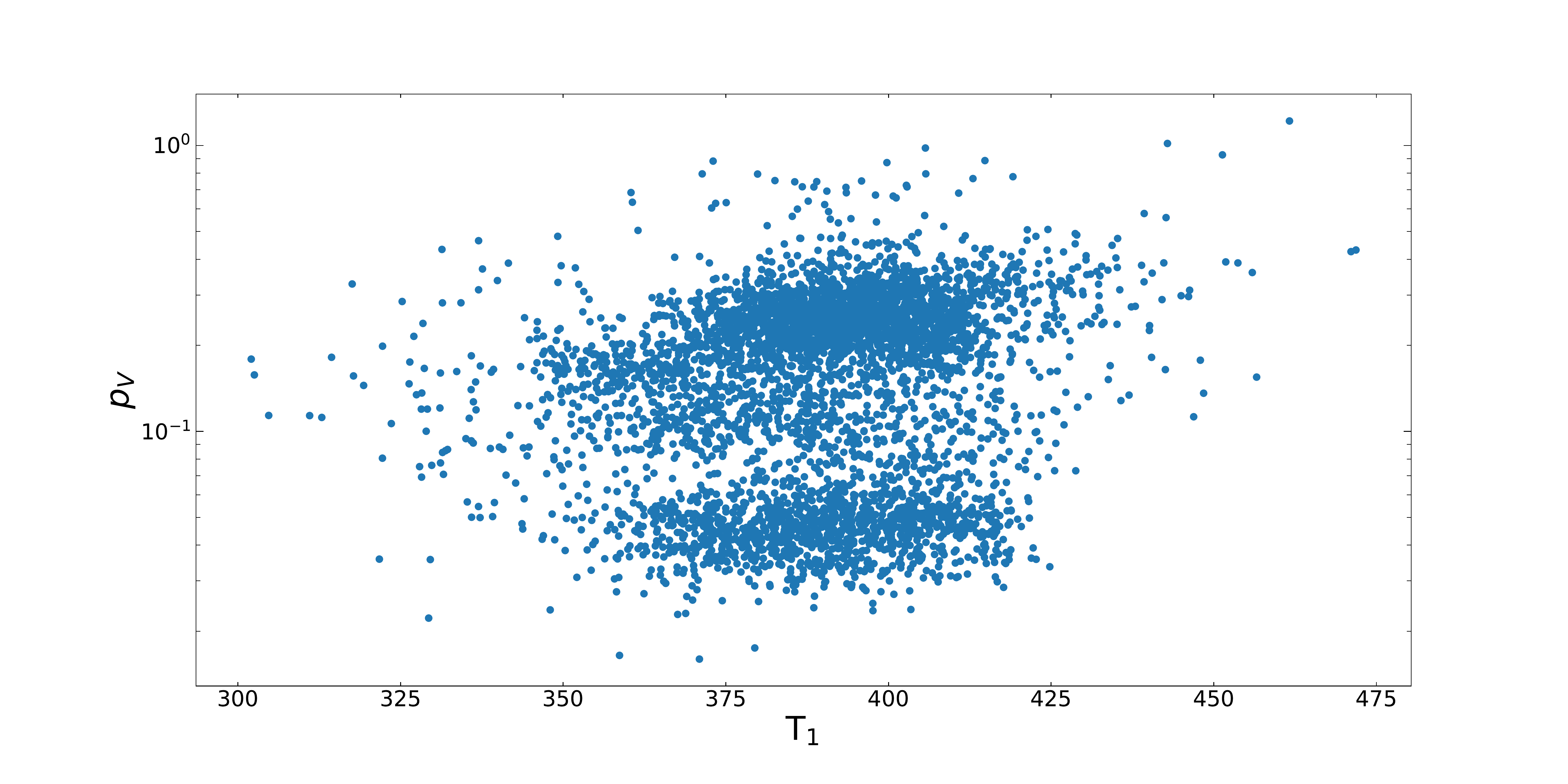}
\caption{Visible-band geometric albedo as a function of pseudo-temperature $T_1$ for the \totnfits asteroid fits presented in this work.}
\label{fig:pv_vs_T1}
\end{center}
  \end{figure}

Figure \ref{fig:pv_vs_T1} shows the visible-band geometric albedo $p_{\rm V}$ as a function of
pseudo-temperature $T_1$ for \totnfits asteroid fits.

\begin{figure}[h]
  \begin{center}
    \includegraphics[width=6.5in]{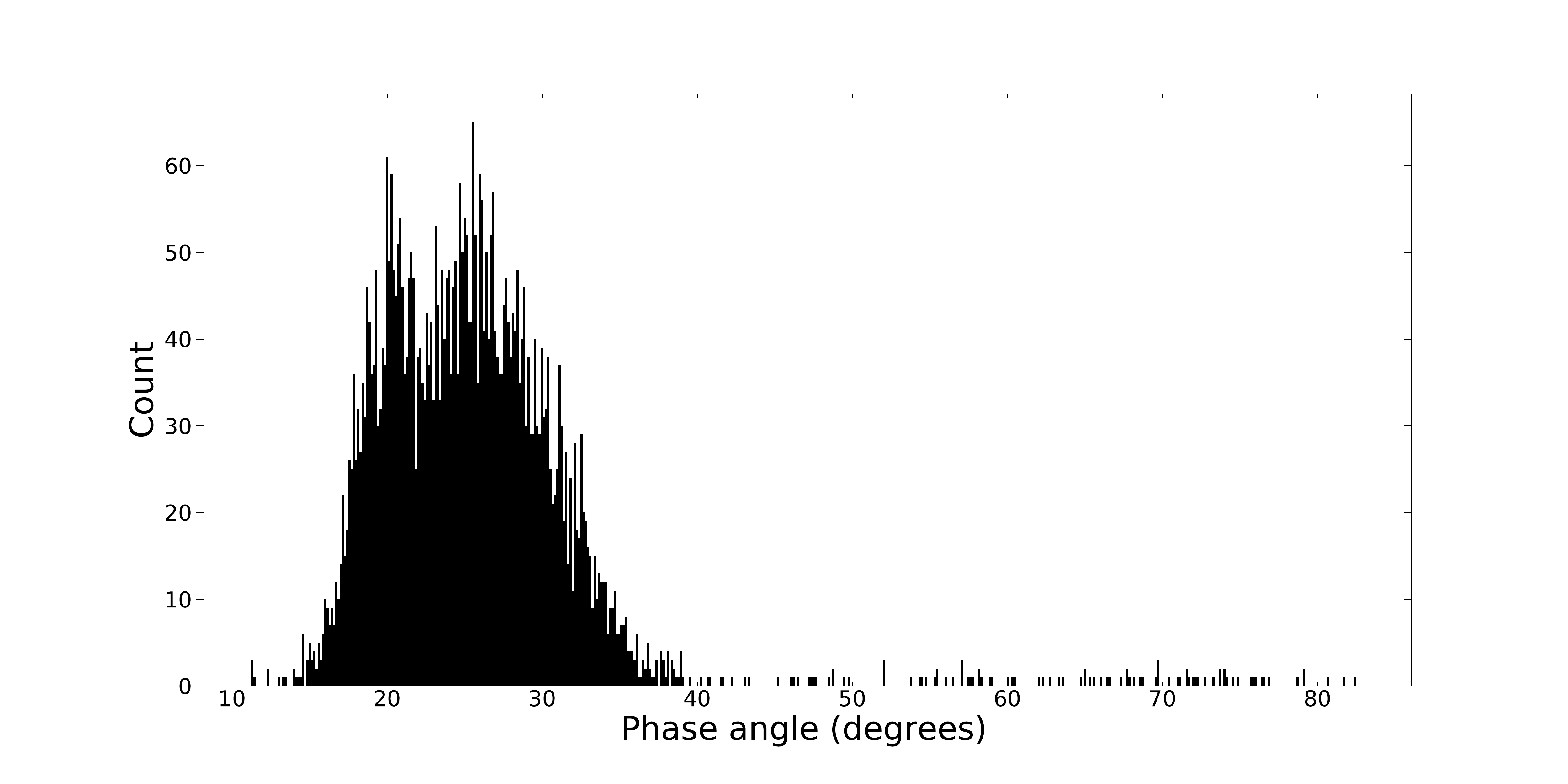}
\caption{Distribution of the median phase angles $\alpha$ for the \totnfits asteroid fits presented in this work.}
\label{fig:phase_angle_hist}
\end{center}
  \end{figure}

Figure \ref{fig:phase_angle_hist} shows the distribution of the median phase angles
$\alpha$ for \totnfits asteroid fits.

\begin{figure}[h]
  \begin{center}
    \includegraphics[width=6.5in]{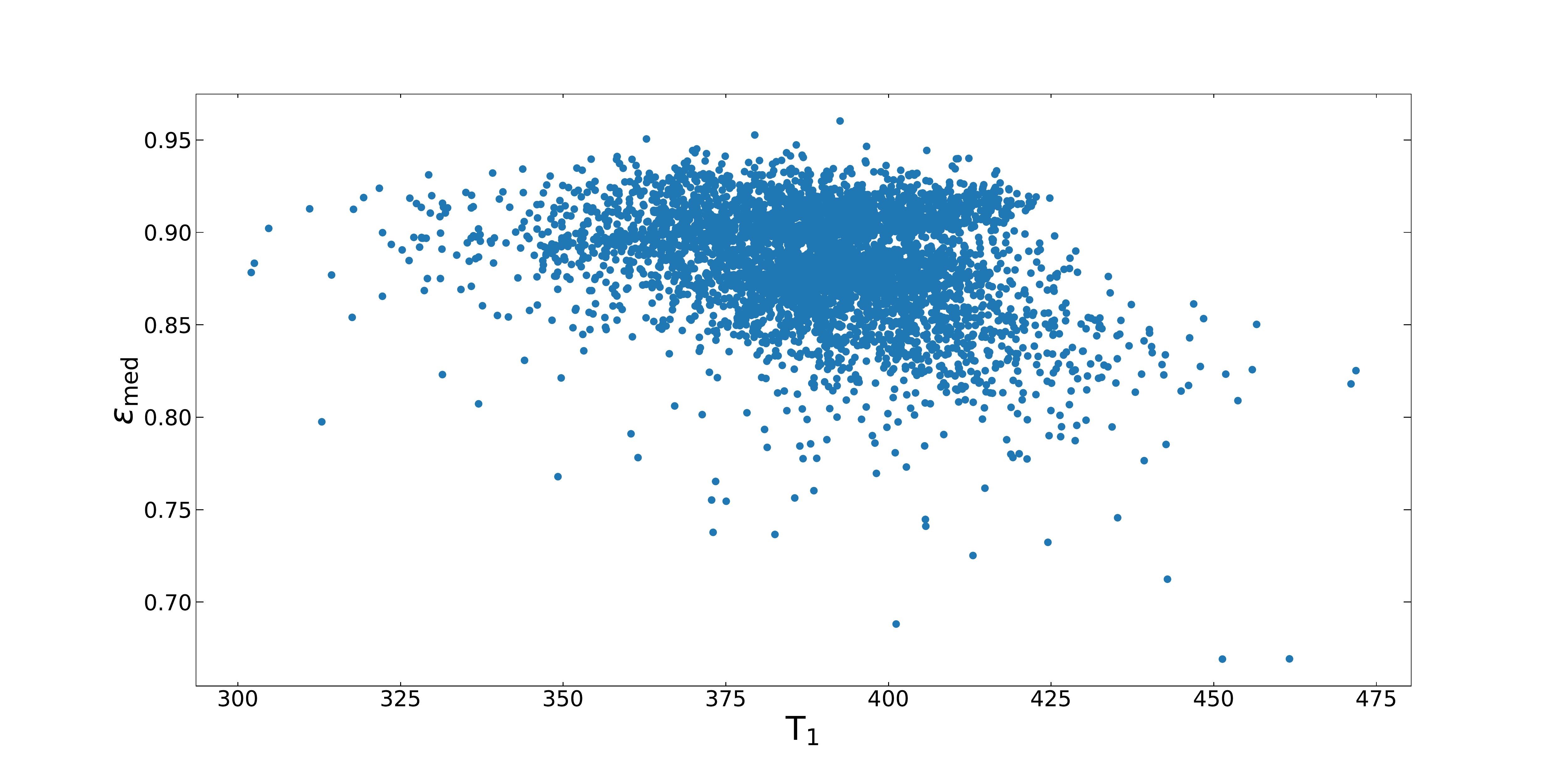}
\caption{Median emissivity value across all four WISE bands as a function of pseudo-temperature $T_1$ for the \totnfits asteroid fits presented in this work.}
\label{fig:eps_med_vs_T1}
\end{center}
  \end{figure}

Figure \ref{fig:eps_med_vs_T1} shows the median emissivity value
$\epsilon_{\rm med}$ across all four WISE bands as a function of
pseudo-temperature $T_1$ for \totnfits asteroid fits.

\begin{figure}[h]
  \begin{center}
    \includegraphics[width=6.5in]{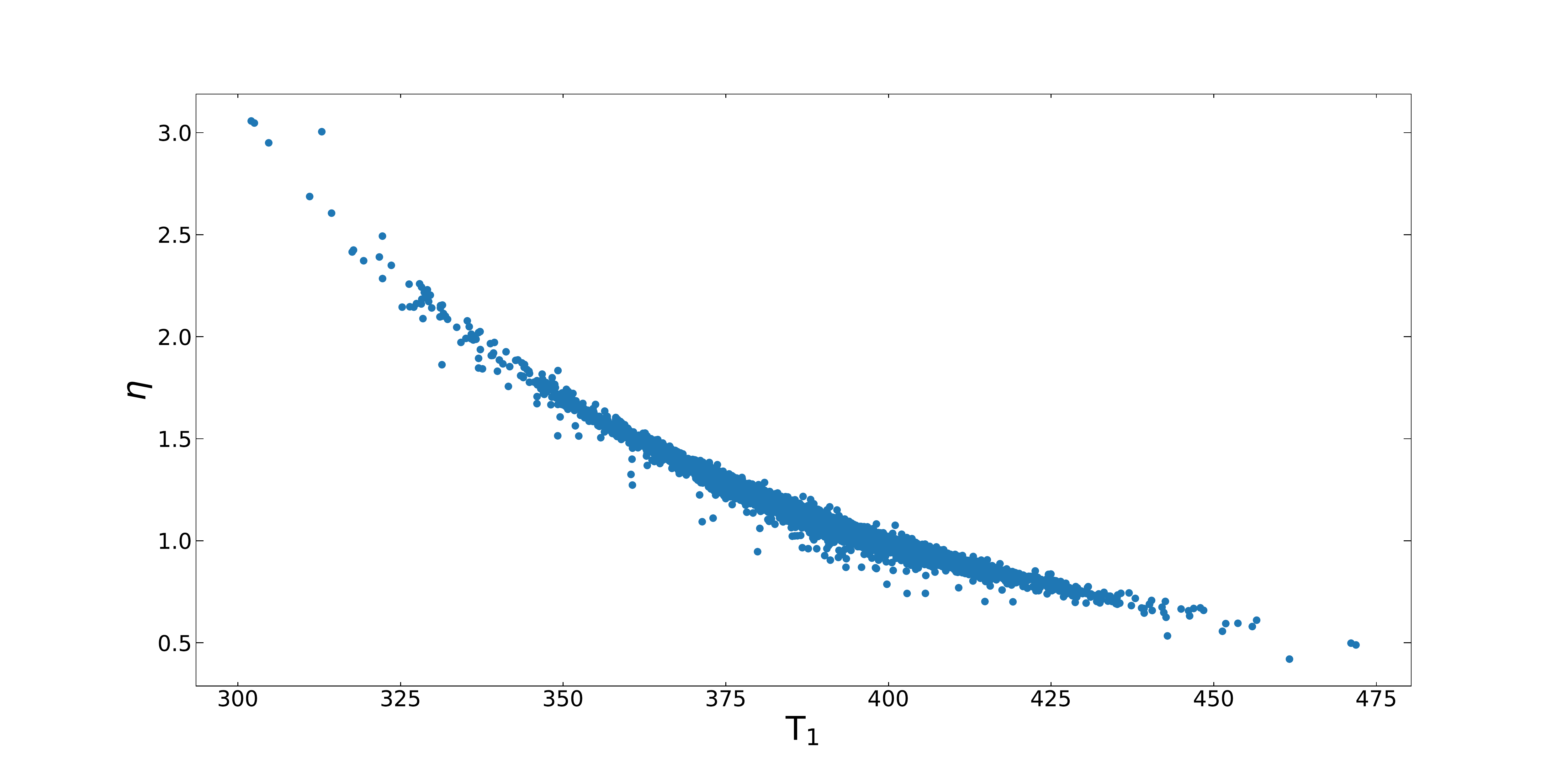}
\caption{Beaming parameter $\eta$ as a function of pseudo-temperature $T_1$ for the \totnfits asteroid fits presented in this work.}
\label{fig:eta_vs_T1_eps}
\end{center}
  \end{figure}

Figure \ref{fig:eta_vs_T1_eps} shows the beaming parameter $\eta$ as a
function of the pseudo-temperature $T_1$ parameter for \totnfits
asteroid fits. The beaming parameter was not a free parameter in the
thermal fits.  Instead, it was calculated using Equation (\ref{eq:T1})
with $\epsilon$ set to the median emissivity value across all four
WISE bands.

\pagebreak
\newpage
\FloatBarrier
\bibliography{neowise}

\begin{thebibliography}{}
\expandafter\ifx\csname natexlab\endcsname\relax\def\natexlab#1{#1}\fi
\providecommand{\url}[1]{\href{#1}{#1}}
\providecommand{\dodoi}[1]{doi:~\href{http://doi.org/#1}{\nolinkurl{#1}}}
\providecommand{\doeprint}[1]{\href{http://ascl.net/#1}{\nolinkurl{http://ascl.net/#1}}}
\providecommand{\doarXiv}[1]{\href{https://arxiv.org/abs/#1}{\nolinkurl{https://arxiv.org/abs/#1}}}

\bibitem[{{Abell} {et~al.}(2015){Abell}, {Barbee}, {Chodas}, {Kawaguchi},
  {Landis}, {Mazanek}, \& {Michel}}]{abel15}
{Abell}, P.~A., {Barbee}, B.~W., {Chodas}, P.~W., {et~al.} 2015, in Asteroids
  IV, ed. P.~{Michel}, F.~E. {DeMeo}, \& W.~F. {Bottke} (Univ. of Arizona
  Press), 855--880, \dodoi{10.2458/azu_uapress_9780816532131-ch043}

\bibitem[{{Benner} {et~al.}(2015){Benner}, {Busch}, {Giorgini}, {Taylor}, \&
  {Margot}}]{benn15}
{Benner}, L.~A.~M., {Busch}, M.~W., {Giorgini}, J.~D., {Taylor}, P.~A., \&
  {Margot}, J.~L. 2015, in {Asteroids IV}, ed. P.~{Michel}, F.~E. {DeMeo}, \&
  W.~F. {Bottke} (Univ. of Arizona Press), 165--182,
  \dodoi{10.2458/azu_uapress_9780816530595-ch009}

\bibitem[{{Bottke} {et~al.}(2015){Bottke}, {Bro{\v{z}}}, {O'Brien}, {Campo
  Bagatin}, {Morbidelli}, \& {Marchi}}]{bott15}
{Bottke}, W.~F., {Bro{\v{z}}}, M., {O'Brien}, D.~P., {et~al.} 2015, in
  Asteroids IV, ed. W.~F. Bottke, A.~Cellino, P.~Paolicchi, \& R.~P. Binzel
  (Univ. of Arizona Press), 701--724,
  \dodoi{10.2458/azu_uapress_9780816532131-ch036}

\bibitem[{{Bottke} {et~al.}(2006){Bottke}, {Vokrouhlick{\'y}}, {Rubincam}, \&
  {Nesvorn{\'y}}}]{bott06}
{Bottke}, Jr., W.~F., {Vokrouhlick{\'y}}, D., {Rubincam}, D.~P., \&
  {Nesvorn{\'y}}, D. 2006, Annual Review of Earth and Planetary Sciences, 34,
  157, \dodoi{10.1146/annurev.earth.34.031405.125154}

\bibitem[{{Bowell} {et~al.}(1989){Bowell}, {Hapke}, {Domingue}, {Lumme},
  {Peltoniemi}, \& {Harris}}]{Bowell89}
{Bowell}, E., {Hapke}, B., {Domingue}, D., {et~al.} 1989, in Asteroids II, ed.
  R.~P. {Binzel}, T.~{Gehrels}, \& M.~S. {Matthews}, 524--556

\bibitem[{Brown {et~al.}(2014)Brown, Jarrett, \& Cluver}]{brown2014}
Brown, M. J.~I., Jarrett, T.~H., \& Cluver, M.~E. 2014, Publications of the
  Astronomical Society of Australia, 31, e049, \dodoi{10.1017/pasa.2014.44}

\bibitem[{Byrd {et~al.}(1995)Byrd, Lu, Nocedal, \& Zhu}]{Byrd1995}
Byrd, R., Lu, P., Nocedal, J., \& Zhu, C. 1995, SIAM Journal of Scientific
  Computing, 16, 1190, \dodoi{10.1137/0916069}

\bibitem[{Coddington {et~al.}(2016)Coddington, Lean, Pilewskie, Snow, \&
  Lindholm}]{LASP}
Coddington, O., Lean, J.~L., Pilewskie, P., Snow, M., \& Lindholm, D. 2016,
  Bulletin of the American Meteorological Society, 97, 1265,
  \dodoi{10.1175/BAMS-D-14-00265.1}

\bibitem[{{Delbo} {et~al.}(2015){Delbo}, {Mueller}, {Emery}, {Rozitis}, \&
  {Capria}}]{delb15}
{Delbo}, M., {Mueller}, M., {Emery}, J.~P., {Rozitis}, B., \& {Capria}, M.~T.
  2015, {Asteroid Thermophysical Modeling}, ed. P.~{Michel}, F.~E. {DeMeo}, \&
  W.~F. {Bottke}, 107--128, \dodoi{10.2458/azu_uapress_9780816532131-ch006}

\bibitem[{{Dermott} {et~al.}(2021){Dermott}, {Li}, {Christou}, {Kehoe},
  {Murray}, \& {Robinson}}]{derm21}
{Dermott}, S.~F., {Li}, D., {Christou}, A.~A., {et~al.} 2021, \mnras, 505,
  1917, \dodoi{10.1093/mnras/stab1390}

\bibitem[{{Farnocchia} {et~al.}(2015){Farnocchia}, {Chesley}, {Milani},
  {Gronchi}, \& {Chodas}}]{farn15}
{Farnocchia}, D., {Chesley}, S.~R., {Milani}, A., {Gronchi}, G.~F., \&
  {Chodas}, P.~W. 2015, in Asteroids IV, ed. P.~{Michel}, F.~E. {DeMeo}, \&
  W.~F. {Bottke} (Univ. of Arizona Press), 815--834,
  \dodoi{10.2458/azu_uapress_9780816532131-ch041}

\bibitem[{Feigelson \& Babu(2012)}]{feigelson2012}
Feigelson, E.~D., \& Babu, G.~J. 2012, Modern statistical methods for
  astronomy: with R applications (Cambridge University Press)

\bibitem[{Greenberg {et~al.}(2020)Greenberg, Margot, Verma, Taylor, \&
  Hodge}]{gree20}
Greenberg, A.~H., Margot, J.-L., Verma, A.~K., Taylor, P.~A., \& Hodge, S.~E.
  2020, The Astronomical Journal, 159, 92, \dodoi{10.3847/1538-3881/ab62a3}

\bibitem[{{Hanu{\v{s}}} {et~al.}(2015){Hanu{\v{s}}}, {Delbo'}, {{\v{D}}urech},
  \& {Al{\'\i}-Lagoa}}]{hanu15}
{Hanu{\v{s}}}, J., {Delbo'}, M., {{\v{D}}urech}, J., \& {Al{\'\i}-Lagoa}, V.
  2015, \icarus, 256, 101, \dodoi{10.1016/j.icarus.2015.04.014}

\bibitem[{{Hapke}(2012)}]{Hapke2012}
{Hapke}, B. 2012, Theory of reflectance and emittance spectroscopy (Cambridge
  university press)

\bibitem[{{Harris}(1998)}]{Harris98}
{Harris}, A.~W. 1998, \icarus, 131, 291,
  \dodoi{https://doi.org/10.1006/icar.1997.5865}

\bibitem[{{Harris} {et~al.}(2015){Harris}, {Boslough}, {Chapman}, {Drube},
  {Michel}, \& {Harris}}]{harr15}
{Harris}, A.~W., {Boslough}, M., {Chapman}, C.~R., {et~al.} 2015, in Asteroids
  IV, ed. P.~{Michel}, F.~E. {DeMeo}, \& W.~F. {Bottke} (Univ. of Arizona
  Press), 835--854, \dodoi{10.2458/azu_uapress_9780816532131-ch042}

\bibitem[{Harris \& Lagerros(2002)}]{harris2002}
Harris, A.~W., \& Lagerros, J.~S. 2002, Asteroids III, 205

\bibitem[{Herald {et~al.}(2019)Herald, Frappa, Gault, Hayamizu, Kerr, Moore, \&
  Giacchini}]{PDSocc}
Herald, D., Frappa, E., Gault, D., {et~al.} 2019, {Asteroid Occultations V3.0},
  {urn:nasa:pds:smallbodiesoccultations::3.0. NASA Planetary Data System},
  \dodoi{10.26033/ap0g-wf63}

\bibitem[{{Herald} {et~al.}(2020){Herald}, {Gault}, {Anderson}, {Dunham},
  {Frappa}, {Hayamizu}, {Kerr}, {Miyashita}, {Moore}, {Pavlov}, {Preston},
  {Talbot}, \& {Timerson}}]{hera20}
{Herald}, D., {Gault}, D., {Anderson}, R., {et~al.} 2020, \mnras,
  \dodoi{10.1093/mnras/staa3077}

\bibitem[{Ivezi{\'c} {et~al.}(2014)Ivezi{\'c}, Connolly, VanderPlas, \&
  Gray}]{ivezic2014}
Ivezi{\'c}, {\v{Z}}., Connolly, A.~J., VanderPlas, J.~T., \& Gray, A. 2014,
  Statistics, data mining, and machine learning in astronomy (Princeton
  University Press)

\bibitem[{Mainzer {et~al.}(2019)Mainzer, Bauer, Cutri, Grav, Kramer, Masiero,
  Sonnett, \& Wright}]{PDSNEOWISE}
Mainzer, A., Bauer, J., Cutri, R., {et~al.} 2019, {NEOWISE Diameters and
  Albedos V2.0}, {urn:nasa:pds:neowise\_diameters\_albedos::2.0. NASA Planetary
  Data System}, \dodoi{10.26033/18S3-2Z54}

\bibitem[{{Mainzer} {et~al.}(2015){Mainzer}, {Usui}, \& {Trilling}}]{main15}
{Mainzer}, A., {Usui}, F., \& {Trilling}, D.~E. 2015, in Asteroids IV, ed.
  P.~{Michel}, F.~E. {DeMeo}, \& W.~F. {Bottke} (Univ. of Arizona Press),
  89--106, \dodoi{10.2458/azu_uapress_9780816532131-ch005}

\bibitem[{{Mainzer} {et~al.}(2011{\natexlab{a}}){Mainzer}, {Bauer}, {Grav},
  {Masiero}, {Cutri}, {Dailey}, {Eisenhardt}, {McMillan}, {Wright}, {Walker},
  {Jedicke}, {Spahr}, {Tholen}, {Alles}, {Beck}, {Brand enburg}, {Conrow},
  {Evans}, {Fowler}, {Jarrett}, {Marsh}, {Masci}, {McCallon}, {Wheelock},
  {Wittman}, {Wyatt}, {DeBaun}, {Elliott}, {Elsbury}, {Gautier}, {Gomillion},
  {Leisawitz}, {Maleszewski}, {Micheli}, \& {Wilkins}}]{main11}
{Mainzer}, A., {Bauer}, J., {Grav}, T., {et~al.} 2011{\natexlab{a}}, \apj, 731,
  53, \dodoi{10.1088/0004-637X/731/1/53}

\bibitem[{{Mainzer} {et~al.}(2011{\natexlab{b}}){Mainzer}, {Grav}, {Masiero},
  {Bauer}, {Wright}, {Cutri}, {McMillan}, {Cohen}, {Ressler}, \&
  {Eisenhardt}}]{main11calibration}
{Mainzer}, A., {Grav}, T., {Masiero}, J., {et~al.} 2011{\natexlab{b}}, \apj,
  736, 100, \dodoi{10.1088/0004-637X/736/2/100}

\bibitem[{{Masiero} {et~al.}(2021){Masiero}, {Wright}, \& {Mainzer}}]{masi21}
{Masiero}, J.~R., {Wright}, E.~L., \& {Mainzer}, A.~K. 2021, The Planetary
  Science Journal, 2, 32, \dodoi{10.3847/PSJ/abda4d}

\bibitem[{{Masiero} {et~al.}(2011){Masiero}, {Mainzer}, {Grav}, {Bauer},
  {Cutri}, {Dailey}, {Eisenhardt}, {McMillan}, {Spahr}, {Skrutskie}, {Tholen},
  {Walker}, {Wright}, {DeBaun}, {Elsbury}, {Gautier}, {Gomillion}, \&
  {Wilkins}}]{masi11}
{Masiero}, J.~R., {Mainzer}, A.~K., {Grav}, T., {et~al.} 2011, \apj, 741, 68,
  \dodoi{10.1088/0004-637X/741/2/68}

\bibitem[{{Moeyens} {et~al.}(2020){Moeyens}, {Myhrvold}, \&
  {Ivezi{\'c}}}]{moey20}
{Moeyens}, J., {Myhrvold}, N., \& {Ivezi{\'c}}, {\v{Z}}. 2020, \icarus, 341,
  113575, \dodoi{10.1016/j.icarus.2019.113575}

\bibitem[{{Myhrvold}(2018{\natexlab{a}})}]{Myhrvold2018ATM}
{Myhrvold}, N. 2018{\natexlab{a}}, \icarus, 303, 91,
  \dodoi{https://doi.org/10.1016/j.icarus.2017.12.024}

\bibitem[{{Myhrvold}(2018{\natexlab{b}})}]{Myhrvold2018Empirical}
---. 2018{\natexlab{b}}, \icarus, 314, 64,
  \dodoi{https://doi.org/10.1016/j.icarus.2018.05.004}

\bibitem[{{Neese}(2010)}]{taxo}
{Neese}, C. 2010, NASA Planetary Data System, EAR

\bibitem[{{Nesvorn{\'y}} {et~al.}(2015){Nesvorn{\'y}}, {Bro{\v{z}}}, \&
  {Carruba}}]{nesv15}
{Nesvorn{\'y}}, D., {Bro{\v{z}}}, M., \& {Carruba}, V. 2015, in Asteroids IV,
  ed. P.~{Michel}, F.~E. {DeMeo}, \& W.~F. {Bottke} (Univ. of Arizona Press),
  297--321, \dodoi{10.2458/azu_uapress_9780816532131-ch016}

\bibitem[{Ostro {et~al.}(2002)Ostro, Hudson, Benner, Giorgini, Magri, Margot,
  \& Nolan}]{ostr02}
Ostro, S.~J., Hudson, R.~S., Benner, L. A.~M., {et~al.} 2002, in Asteroids III,
  ed. W.~F. Bottke, A.~Cellino, P.~Paolicchi, \& R.~P. Binzel (Univ. of Arizona
  Press), 151--168

\bibitem[{Pravec {et~al.}(2012)Pravec, Harris, Kušnirák, Galád, \&
  Hornoch}]{prav12}
Pravec, P., Harris, A.~W., Kušnirák, P., Galád, A., \& Hornoch, K. 2012,
  Icarus, 221, 365 , \dodoi{https://doi.org/10.1016/j.icarus.2012.07.026}

\bibitem[{Ratkowsky(1983)}]{ratkowsky1983}
Ratkowsky, D. 1983

\bibitem[{{Scheeres} {et~al.}(2015){Scheeres}, {Britt}, {Carry}, \&
  {Holsapple}}]{sche15}
{Scheeres}, D.~J., {Britt}, D., {Carry}, B., \& {Holsapple}, K.~A. 2015, in
  Asteroids IV, ed. P.~{Michel}, F.~E. {DeMeo}, \& W.~F. {Bottke} (Univ. of
  Arizona Press), 745--766, \dodoi{10.2458/azu_uapress_9780816532131-ch038}

\bibitem[{Vereš {et~al.}(2015)Vereš, Jedicke, Fitzsimmons, Denneau, Granvik,
  Bolin, Chastel, Wainscoat, Burgett, Chambers, Flewelling, Kaiser, Magnier,
  Morgan, Price, Tonry, \& Waters}]{veres2015}
Vereš, P., Jedicke, R., Fitzsimmons, A., {et~al.} 2015, Icarus, 261, 34 ,
  \dodoi{https://doi.org/10.1016/j.icarus.2015.08.007}

\bibitem[{Warner, B.D. and Harris, A.W. and Pravec, P.(2021)}]{lcdb}
Warner, B.D. and Harris, A.W. and Pravec, P. 2021, Asteroid Lightcurve Data
  Base (LCDB) Bundle V4.0., NASA Planetary Data System,
  \dodoi{10.26033/j3xc-3359}

\bibitem[{{Wright} {et~al.}(2018){Wright}, {Mainzer}, {Masiero}, {Grav},
  {Cutri}, \& {Bauer}}]{wrig18}
{Wright}, E., {Mainzer}, A., {Masiero}, J., {et~al.} 2018, arXiv e-prints,
  arXiv:1811.01454.
\newblock \doarXiv{1811.01454}

\bibitem[{{Wright}(2019)}]{wrig19aas}
{Wright}, E.~L. 2019, in American Astronomical Society Meeting Abstracts, Vol.
  233, American Astronomical Society Meeting Abstracts \#233, 263.04

\bibitem[{{Wright} {et~al.}(2010){Wright}, {Eisenhardt}, {Mainzer}, {Ressler},
  {Cutri}, {Jarrett}, {Kirkpatrick}, {Padgett}, {McMillan}, {Skrutskie},
  {Stanford}, {Cohen}, {Walker}, {Mather}, {Leisawitz}, {Gautier}, {McLean},
  {Benford}, {Lonsdale}, {Blain}, {Mendez}, {Irace}, {Duval}, {Liu}, {Royer},
  {Heinrichsen}, {Howard}, {Shannon}, {Kendall}, {Walsh}, {Larsen}, {Cardon},
  {Schick}, {Schwalm}, {Abid}, {Fabinsky}, {Naes}, \& {Tsai}}]{Wright2010}
{Wright}, E.~L., {Eisenhardt}, P. R.~M., {Mainzer}, A.~K., {et~al.} 2010, \aj,
  140, 1868, \dodoi{10.1088/0004-6256/140/6/1868}

\end{thebibliography}
\vfill

\end{document}